\documentclass[a4paper]{jpconf}
\usepackage{graphicx}
\usepackage{amsmath, amsthm, amssymb}
\usepackage{bbold}
\usepackage{epsfig}
\usepackage[latin1]{inputenc}
\usepackage{cite}
\usepackage{array}
\usepackage{url}
\usepackage{mathtools}
\usepackage{amsfonts}
\usepackage{algpseudocode} 
\usepackage{float}
\usepackage{paralist}
\usepackage{xspace}
\usepackage{LocalDefinitions}
\begin{document}
\title{\scampi: a robust approximate message-passing framework for compressive imaging}
\author{Jean Barbier$^\dag$, Eric W. Tramel$^{\ddag}$, and Florent Krzakala$^{\ddag}$}
\address{$\dag$ Laboratoire de Th\'eorie des Communications, 
                Ecole Polytechnique F\'ed\'erale de Lausanne, 
                Facult\'e Informatique et Communications, 
                CH-1015, Suisse.}
\address{$\ddag$ Laboratoire de Physique Statistique CNRS UMR 8550, 
                 Universit\'e P. et M. Curie et Ecole Normale Sup\'erieure, 
                 24 rue Lhomond, 75005 Paris, France.}
\ead{jean.barbier@epfl.ch, eric.tramel@ens.fr, florent.krzakala@ens.fr}
\begin{abstract} 
Reconstruction of images from noisy linear measurements is a core problem in image 
processing, for which convex optimization methods based on total variation (TV)
minimization have been the long-standing state-of-the-art. We present an 
alternative probabilistic reconstruction procedure based on approximate message-passing, \scampi, which operates in the compressive regime, where the inverse
imaging problem is underdetermined. While the proposed method 
is related to the recently proposed GrAMPA algorithm of 
Borgerding, Schniter, and Rangan, 
we further develop the probabilistic approach to compressive imaging by introducing
an expectation-maximizaiton learning of model parameters, making the \scampi 
robust to model uncertainties. Additionally, our numerical experiments indicate
that \scampi can provide reconstruction performance superior to both GrAMPA as 
well as convex approaches to TV reconstruction. 
Finally, through exhaustive best-case experiments, we show that in many cases
the maximal performance of both \scampi and convex TV can be quite close, even
though the approaches are \emph{a prori} distinct. The theoretical reasons for 
this correspondence remain an open question. Nevertheless, the proposed algorithm
remains more practical, as it requires far less parameter tuning to perform 
optimally.
\end{abstract}
\section{Introduction and problem setting}
\label{sec:into}
% !TEX root = ../main.tex
Over the past decade, the study of compressed sensing (CS) 
\cite{Candes:2005um,CandesTao:06,CandesRombergTao06,Donoho:06}
has blossomed into a
large field of active research in signal processing. Of particular interest in this
field is the development of sparse signal reconstruction algorithms from noisy, linear 
measurements. While the development of generic algorithms for this problem is well
advanced, there still remains much work to be done on the development of reconstruction
algorithms that exploit the statistics of particular signal classes. In this work,
we are interested in the class of \emph{natural images} which has its application within
compressive imaging \cite{DDT2008}. 
Natural images are considered as a class of \twoD signals which
posses a \emph{piecewise-continuity}, exhibiting large regions of the signal support which are constant or slowly varying interspersed with singularities of rapid transitions,
or \emph{edges}. In this work we will see how this property of piecewise-continuity can
be exploited to accurately reconstruct compressively sampled images, subject to varying
degrees of measurement degradation.

For the present study, we consider the reconstruction of a vectorized image signal
$\plantedsignal\in\mathbb{R}^{N}$ composed of $N = L^2$ pixels from linear 
measurements $\measures\in\mathbb{R}^{M}$ which represent the observations of the
signal projected by the measurement matrix $\F \in \mathbb{R}^\ofdim{M}{N}$. 
The measurements are corrupted by an additive white Gaussian noise (AWGN)
which is zero mean and of variance $\Delta^*$,
$\noise \sim \mathcal{N}(\noise;\mathbf{0}_\ofdim{M}{1},\Delta^* \mathbf{I}_\ofdim{M}{M})$ where $\mathbf{I}_\ofdim{a}{a}$ is the identity matrix of 
size $a\times a$ and $\boldsymbol{0}_\ofdim{a}{b}$ is a matrix of size $a\times b$ whose elements are all 0. With these definitions, the undersampled observation procedure can be written as
\begin{equation}
  \measures = \F\plantedsignal + \noise 
            \quad \Leftrightarrow \quad
  \measure = \plantedprojection + \noiseu,
  \label{eqIntro:AWGNCS} 
\end{equation}
where $\langle\cdot,\cdot\rangle$ represents an inner-product and $\F_{\mu}$ is 
the $\mu^{\rm th}$ row of $\F$.
The fundamental theory of CS asserts that with a properly chosen $\F$, usually random, one can accurately recover $\plantedsignal$ in the under-determined 
setting of $M<N$, by introducing \emph{a priori} knowledge of $\plantedsignal$, namely,
sparsity. A sparse signal is one which, either in the cardinal basis or some other
transform basis, posses only a few non-zero components. To be more precise, in an \emph{analysis model} \cite{Nam201330}, it is assumed that a linear transform $\Psi\plantedsignal$ of the signal $\plantedsignal$ to be reconstructed is sparse, while $\plantedsignal$ is generally not. In this case, $\plantedsignal$ is sometimes refered as \emph{cosparse}. Complementary to this definition is the \emph{synthesis model}, in which it is directly the signal $\plantedsignal$ of interest that is assumed to be sparse, i.e. $\Psi$ is the identity.
The generic reconstruction task, 
in the case of AWGN corrupted observations, can thus be written as a Lasso \cite{Tib1996} problem,
\begin{equation}
  \hat{\signal}_{\ell_1} 
    = \underset{\signal\in\mathbb{R}^N}{\text{argmin}} \
      || \measures - \F\signal ||_2^2 + \lambda ||\Psi \signal||_1,
\end{equation}
where $\lambda$ is a regularization parameter whose optimal selection is generally
unknown \emph{a posteriori}, and 
the $\ell_p$ norm is defined as 
$||\mathbf{x}||_p \defas \paren{\sum_i |x_i|^p}^{1/p}$.
While one might entertain the
use of generic sparse reconstruction techniques for images by recovering sparse 
coefficients in, say, the wavelet basis, the use of gradient variation minimization
has remained state-of-the-art for image reconstruction in CS. Specifically, we are
interested in the application of total variation (TV) minimization \cite{CandesRombergTao06}
for image reconstruction.  
In this approach, one biases the optimization towards solutions which exhibit a 
sparse \twoD discrete gradient. In the case of anisotropic TV minimization for
AWGN corrupted observations, the reconstruction problem can be written as
\begin{equation}
  \hat{\signal}_{\rm TV} 
    = \underset{\signal\in\mathbb{R}^N}{\text{argmin}} \
      || \measures - \F\signal ||_2^2 + \lambda \sum_{i=1}^{N} \left(|\Gh_{s_i}| + |\Gv_{s_i}|\right),
\end{equation}
where $\Gh_{s_i}$ and $\Gv_{s_i}$ are the horizontal and vertical discrete gradients of $\signal$, respectively, at pixel index $i$. This problem has been well studied from the perspective of convex
optimization, with many recent developments focused on improving the rate-of-convergence
and computational efficiency of algorithms to solve this problem \cite{CP2011}. One robust
and efficient reconstruction algorithm which we utilize in this study is the TV-AL3 of \cite{Li2009}, based on the augmented Lagrangian approach.

Rather than studying the TV minimization problem as a convex one, we will instead adopt
a probabilistic framework that mimics the original TV problem, in the 
sense that it will try to output piecewise-continuous 
solutions of the system \eqref{eqIntro:AWGNCS}. 
The use of statistical approaches to TV minimization is not novel to this work. For instance,
the \oneD TV problem, and its associated phase transitions,
was studied in \cite{DJM2013} and a reconstruction method based on
approximate message-passing (AMP) was first proposed.
In a related vein, there also exist applications of AMP to image representation via non-local
means \cite{DBLP:journals/corr/MetzlerMB14} as well as 
AMP in conjunction with an amplitude-scale-invariant Bayes estimator 
\cite{DBLP:journals/corr/TanMB14}. 
Another recent work for \oneD TV proposes a joint prior defined over nearest-neighbors 
components, characterizing the gradient with a spike-and-slab prior 
within AMP \cite{KJL2014} (SS-AMP). 
Finally, one of the most exciting applications of the generalized AMP \cite{Rangan10b} to \twoD signals has been
the work of \cite{BP2013}, which proposed the use of a cosparse analysis model. This approach, known as GrAMPA, is also general in the sense of non-AWGN i.i.d observation channels and can be used with any sparsifying basis $\Psi$. 
GrAMPA was shown to be a state-of-the-art approach for image reconstruction in the
benchmark tests of \cite{KJL2014,BP2013}.
Motivated by the results of both SS-AMP and GrAMPA, we present a new AMP-based 
algorithm for \twoD CS image reconstruction particularly adapted to natural images.

Similar to GrAMPA, the proposed method will also work with the cosparse analysis model. Furthermore, we utilize
the sparse non-informative parameter estimator (SNIPE) prior proposed in \cite{BP2013} for auxiliary gradient variables that will be introduced in the model. We term this approach 
``SNIPE-based cosparse analysis AMP for imaging'' (\scampi).
We will detail the novel modifications of our approach with respect to GrAMPA in the sequel, but they can be summarized as 
\begin{inparaenum}[\itshape i\upshape)]
  \item an expectation-maximization learning of the noise variance, 
        which makes the algorithm more robust to channel uncertainty, and
  \item the introduction of auxiliary 
        variables that allow one to relax the constraints over the differences between 
        neighboring pixels.
\end{inparaenum}
In terms of statistical physics, our proposed approach can be seen as a 
``finite temperature'' version of GrAMPA. This relaxation improves reconstruction
accuracy for natural images, where algorithms must be more robust to violations of
the piecewise-continuous assumption. 

We also show that in the limit of oracle parameter selection, the performances of TV-AL3 and \scampi are comparable, and in the low-measurement regime, the differences are almost negligible. This limiting performance raises a number of interesting questions about the nature and maximal performance of TV-based CS reconstruction of natural images.
\section{Proposed model and algorithm: \scampi}
\label{sec:scampi}
% !TEX root = ../main.tex
\subsection{The cosparse TV analysis model with dual variables}
Let us define the set of all edges between neighboring pixels in the image as $E \defas \bra{\edge = (i,j):(i,j) \in \{1,\dots,N\}\times\{1,\dots,N\}, s_i \ \text{neighbor of} \ s_j}$. 
Pixel neighborhoods can be chosen arbitrarily, but to match the TV cost function,
we will limit ourselves to neighborhoods corresponding to a 4-connected lattice, placing
edges between horizontally and vertically co-located pixels. Next, we introduce a set of
auxiliary, or \emph{dual}, variables which exist on the edges which describe the differences
between individual pixels,
\begin{equation}
  \dualvars \defas [(s_i - s_j) : (i,j) = \edge \in E].
\end{equation}
We now define the augmented signal
$\augsignal \defas [\signal, \dualvars]^{\intercal}$ as the concatenation of the image 
representation and the vector of dual variables. We can now write out an augmented system for \eqref{eqIntro:AWGNCS} which introduces the dual variables
\begin{align}
  \left[ \begin{array}{c}
           \measures_\ofdim{M}{1}\\
           \boldsymbol{0}_\ofdim{|E|}{1}
         \end{array}  
  \right]
  &= 
  \left[ \begin{array}{cc}
           \F_\ofdim{M}{N}        &   \boldsymbol{0}_\ofdim{M}{|E|} \\
           \fdmatrix_\ofdim{|E|}{N} & -\mathbf{I}_\ofdim{|E|}{|E|} 
         \end{array}  
  \right]
  \times
  \left[ \begin{array}{c}
          \signal_\ofdim{N}{1}\\
          \dualvars_\ofdim{|E|}{1} 
         \end{array}  
  \right]
  +
  \left[ \begin{array}{c}
           \noise_\ofdim{M}{1}\\
           \dualnoise_\ofdim{|E|}{1} 
          \end{array}  
  \right],
\label{eq:hatSyst}
\end{align}
where $\F$ is the original measurement matrix and the dimensions of each vectors and matrices have been indicated in subscript 
for clarity. The entire system can be written and solved in the reduced form: $\augmeasures = \augsystem\augsignal + \augnoise$.

$\fdmatrix$ is a \twoD finite difference operator, 
i.e. the concatenation of two matrices matrices which calculate both the horizontal
and vertical gradients. As stated earlier, we consider for each pixel its four closest neighbors, 
thus $\fdmatrix \defas[\fdmatrixRight, \fdmatrixDown]^\intercal$ is the concatenation of 
$\fdmatrixRight$ which is made of zeros everywhere except for a main
diagonal of $1$'s, and a single $-1$ on each row, in a column corresponding to the
\emph{right} neighbor of the pixel at the index represented by the $1$ on the main diagonal. 
The matrix $\fdmatrixDown$ is constructed in similar fashion, but for the \emph{down} nieghbor.
This choice of right/down differencing, versus left/up, is arbitrary. As long as $\fdmatrix$
is constructed consistently, all pairwise interactions are encoded consistently.
Finally, the vector $\dualnoise$ is an error that can be thought of as the 
degree of correspondence
between $\dualvars$ and the true discrete \twoD gradient of $\signal$. We will discuss
this point more throughly in the next section.
As a whole, we refer to this augmented system as a {\it{cosparse TV analysis model with dual variables}}.

\subsection{Probabilistic model for TV reconstruction}
We now present a Bayesian framework for the TV problem. 
The above construction allows one to easily associate 
a factorized prior $\augprior$ to $\augsignal$, a key assumption for the AMP algorithm 
\cite{DonohoMaleki09,BayatiMontanari10,DonohoMaleki10,KMS2012b,KrzakalaMezard12,JavanmardMontanari12}.
While some arbitrary \iid distribution can be assigned to the elements of $\signal$,
we will assume a SNIPE 
prior $\snipeprior$ \cite{BP2013} for $\dualvars$. This prior
is defined as the limiting distribution of a spike-and-slab distribution with infinite variance, 
\begin{equation} 
  \snipeprior(x;\omega) \defas
    \lim_{\sigma\to\infty} \frac{\rho(\sigma,\omega)}{\sigma}\mathcal{N}(x/\sigma;0,\sigma^2)
                          + (1-\rho(\sigma,\omega)) \delta(x),
\end{equation}
for which, under a proper scaling of $\rho(\sigma,\omega)$, has $\omega$ as the only 
free parameter of the prior. For further details on SNIPE, we direct the reader to \cite{BP2013}.
By using $\snipeprior$ to bias the elements of $\dualvars$ during inference, 
we place more probabilistic weight on images whose finite
differences are sparse, according to $\omega$. 
  
If we consider that the elements of $\signal$ obey a uniform distribution, and if we enforce that 
the elements of the concatenated signal $\augsignal$ are ordered such that the 
first $N$ elements correspond to pixel values and the last $|E|$ elements correspond to $\dualvars$,
then the factorized prior for $\augsignal$ can be writen as
\begin{equation}
  \augprior(\augsignal) 
    \propto \prod_{i=1}^{N+|E|} P_0^i(x_i) 
    =       \prod_{i=1}^{N+|E|} \left[  \mathbb{I}(i\le N)\mathcal{U}(x_i) 
                                      + \mathbb{I}(i>N) \snipeprior(x_i;\omega) \right],
\label{eq:facPrior}
\end{equation}
where $\mathcal{U}$ is a uniform distribution over some domain of $\mathbb{R}$
and $\mathbb{I}(\cdot)$ is the indicator function. 
%%% I think that this invites too many questions and we don't have experiments for it...
% One could 
% think about enforcing positivity by taking $\mathbb{R}^+$ but it complicates the equations and 
% empirically does not improve the reconstruction performances.
%%%

With the prior defined for $\augsignal$, we would like to turn our attention to the 
posterior distribution of $\augsignal$ given $\measures$, however, first, we must take a 
moment to consider the likelihood induced by the distributions of the
error terms $\noise$ and $\dualnoise$.
We diverge from the GrAMPA approach, by the introduction of the ``finite temperature''
noted in the previous section. In order to mimic the GrAMPA approach thanks to the cosparse TV analysis model with dual variables (\ref{eq:hatSyst}), one would set $\dualnoise = \boldsymbol{0}_\ofdim{|E|}{1}$, creating a model for which the elements of $\dualvars$ are exactly the discrete gradient of $\signal$. Indeed, this is an intuitive construction. However, imposing such a strict 
condition might impede the reconstruction. The fist key feature of our approach is that instead, we propose that the hard 
linear constraints $\{d_{\edge}-(s_i-s_j)=0:(i,j)=\edge\in E\}$ be ``relaxed'' by taking 
$\dualnoise \sim \mathcal{N}(\dualnoise;\boldsymbol{0}_\ofdim{|E|}{1},T~\mathbf{I}_\ofdim{|E|}{|E|})$. 
Here, the amount of relaxation is determined by the temperature-like variance parameter, $T$.
In the frozen zero-temperature limit, $T\to0$, one recovers GrAMPA, up to the nuances of 
implementation. However, for finite temperature, some slack is introduced into the problem, allowing the dual variables to fluctuate some distance away from the true gradient, a helpful feature to avoid poor local solutions.

\begin{figure}[!t]
\centering
\includegraphics[width=0.5\textwidth]{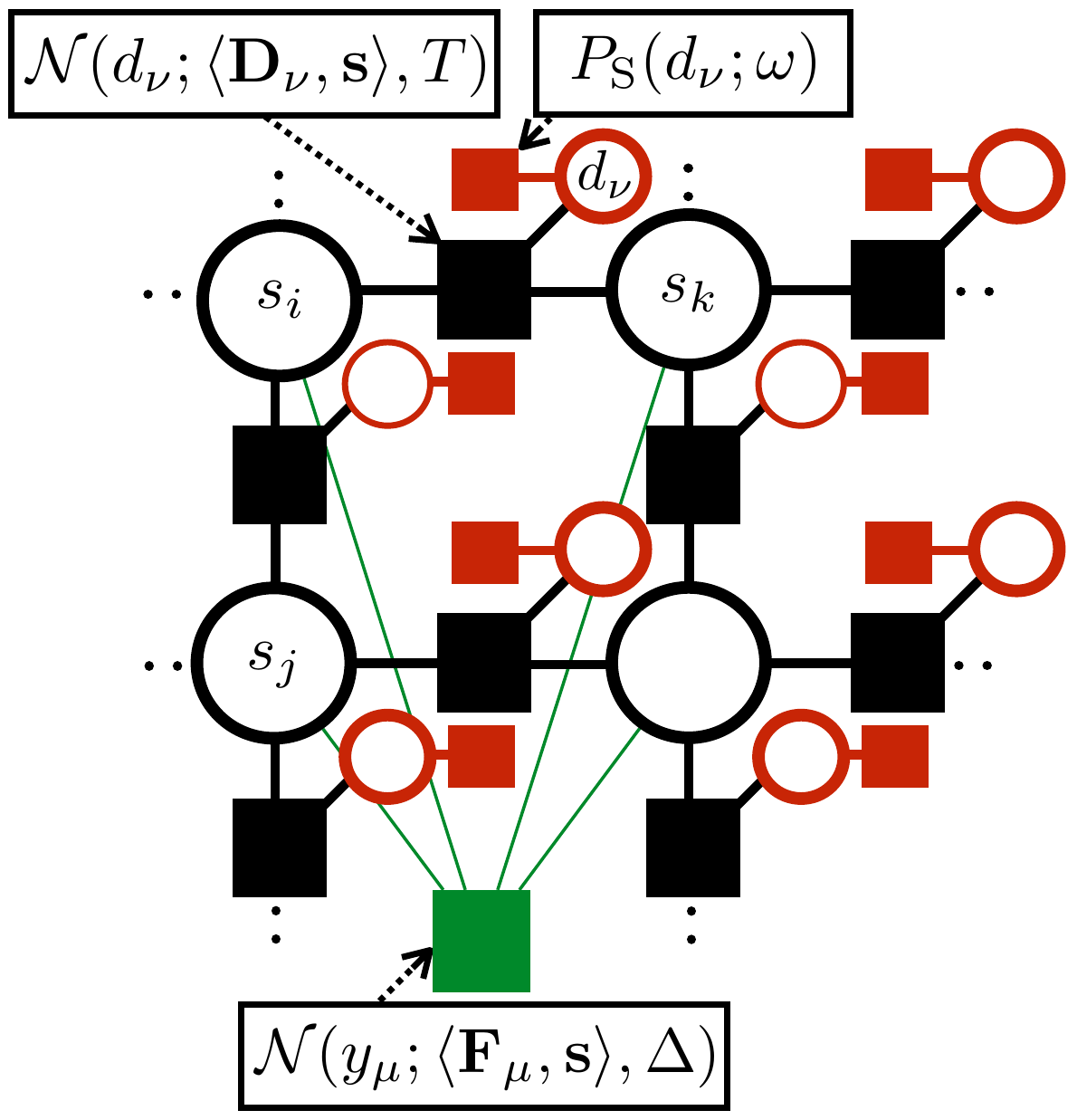}
\caption{
  Factor graph associated to the probability measure \eqref{eq:pampd} sampled by the \scampi 
  algorithm for the image reconstruction. The large black circles are the pixels of $\signal$, 
  while the smaller red ones are the dual variables, $\dualvars$. 
  The squares represent the factors: the green ones 
  connected to all the pixel variables are the ``measurement'' factors 
  enforcing $\signal$ to 
  verify the linear measurements $\measures$, up to Gaussian fluctuations of variance 
  $\Delta$. The red factors represent the factorizable prior we assume about the 
  dual variables, which is the SNIPE prior $\snipeprior$. Finally, the black 
  factors are enforcing consistency between $\signal$ and $\dualvars$, as weighted by $T$. 
  To avoid double counting, for each pixel, only two out of the four possible neighbor interactions
  are taken into account.}
\label{fig:factorGraph}
\end{figure}

Finally, we can now write the posterior distribution for our TV model (\ref{eq:hatSyst}) as
\begin{equation}
\scampiposterior(\augsignal = [\signal, \dualvars] | \augmeasures = [\measures, \boldsymbol{0}])
  = \frac{1}{\Z}   
    \prod_{\mu=1}^M \mathcal{N}\left(\measure;\projection, \Delta\right) 
    \prod_{\nu=1}^{|E|} 
      \left[\mathcal{N}\left(\dualvar;\gradient, T\right)\snipeprior(d_\nu;\omega) \right],
\label{eq:pampd}
\end{equation}
where the normalization $\Z \defas Z(\measures, \F, \fdmatrix, \Delta, T, \omega)$,
also known as the partition function in physics, is a function of the problem instance, 
and is intractable to calculate, in practice. If known, the noise variance associated with the 
measurements in the posterior 
should match the true noise variance, $\Delta = \Delta^{*}$. If $\Delta^{*}$ is unknown, it may
be estimated,  
as we will see in Sec. \ref{sec:learnnoise}.
Rather than attempting to calculate this
posterior exactly, the message-passing approach we discuss, here, attempt to \emph{approximate}
this posterior via information transfer between nodes on the factor graph representation 
of the posterior. Fig.~\ref{fig:factorGraph} gives the factor graph 
representation of the proposed \scampi posterior. The factor graph associated with 
the posterior sampled by GrAMPA is the same, but lacks the red variables and factors. Also, the black factors would be of the 
form $\snipeprior(\gradient;\omega)$ instead of the proposed Gaussian densities, i.e. 
$\grampaposterior(\signal) \propto \prod_{\mu=1}^M \mathcal{N}\left(\measure;\projection, \Delta\right) \prod_{\nu=1}^{|E|} \snipeprior(\gradient;\omega)$. 

Lastly, we would like to mention that one can also consider a \twoD extension of SS-AMP
without appealing to the augmented analysis model (\ref{eq:hatSyst}). In this case, the posterior becomes
$\ssampposterior(\signal) \propto \prod_{\mu=1}^M \mathcal{N}\left(\measure;\projection, \Delta\right) \prod_{\nu=1}^{|E|} [\rho \mathcal{N}(\gradient;0, T) + (1-\rho) \delta(\gradient)]$.
However, empirical results indicate that this algorithm suffers from strong convergence issues.
Additionally, the exploding complexity, and quantity, of the resulting message-passing equations 
create too many numerical issues for a practical implementation at scale.
\subsection{AMP for the cosparse analysis model with dual variables: \scampi}
The \scampi algorithm, detailed in Fig.~\ref{algoCh1:AMP}, is  
canonical AMP applied 
to the cosparse analysis model with dual variables \eqref{eq:hatSyst}, with a few added
features we detail in the sequel. \scampi performs a minimum mean square error estimation of $\augsignal$ by sampling from an iteratively computed approximation of the distribution 
\eqref{eq:pampd}. The full derivation of AMP in similar notations can be found in \cite{KrzakalaMezard12,PhDjeanBarbier}. Additionally, we define the augmented vector of per-factor noise variances as
$\mathring{\Delta}_\mu\defas \Delta \mathbb{I}(\mu \le M) + T \mathbb{I}(\mu > M)$. 

\begin{figure}[!t]
\begin{algorithmic}[1]
    \State $t\gets 0$
    \State $\convg \gets 1 + \epsilon$
    \While{$t<t_{max} \ \textbf
    {and} \ \convg >\epsilon$} 
    \State $\tilde\Theta\NextIter_\mu \gets \sum_{i=1}^{N+|E|}\mathring{F}_{\mu i}^2 v_i\ThisIter$
    \State $w\NextIter_\mu \gets \damp w\ThisIter_\mu + (1-\damp)\(\sum_{i=1}^{N+|E|} \mathring{F}_{\mu i}a_i\ThisIter - \tilde\Theta\NextIter_\mu\frac{\augmeasure-w\ThisIter_\mu}{\augnoisevaru\ThisIter + \Theta\ThisIter_\mu}\)$
    \State $\Theta\NextIter_\mu \gets \damp \Theta\ThisIter_\mu + (1-\damp)\tilde\Theta\NextIter_\mu$
    \State $\Sigma\NextIter_i \gets \paren{\sum_{\mu=1}^{M+|E|}\frac{\mathring{F}_{\mu i}^2}{\augnoisevaru\ThisIter + \Theta_\mu\NextIter}}^{-\half}$
    \State $R\NextIter_i \gets a\ThisIter_i + \paren{\Sigma\NextIter_i}^2 \sum_{\mu=1}^{M+|E|} \mathring{F}_{\mu i}\frac{\mathring y_\mu - w\NextIter_\mu}{\augnoisevaru\ThisIter + \Theta\NextIter_\mu}$
    \State $v\NextIter_i \gets \fc\left(\paren{\Sigma_{i}\NextIter}^2,R_{i}\NextIter;\omega\right)$
    \State $a\NextIter_i \gets \fa\left(\paren{\Sigma_{i}\NextIter}^2,R_{i}\NextIter;\omega\right)$
    \State Update all $\augnoisevaru\NextIter$ according to \eqref{eq:noiseLearn}
    \State $\convg \gets \frac{1}{N}||\ba\NextIter - \ba\ThisIter||_2^2$
    \State $t \gets t+1$
    \EndWhile
    \State \textbf{return} $\ba\ThisIter$
\end{algorithmic}
\caption{%
  The \scampi algorithm with damping (controlled by $0\le\damp<1$) for the co-sparse model with 
  dual variables. The convergence tolerance is given by $\epsilon$ and $t_{max}$ is the maximum number of 
  iterations. A suitable initialization for the posterior mean and variance are, respectively,
  $a_i\InitIter = \mathbb{E}_{P_0^i}(x_i)$ and
  $v_i\InitIter=\txt{Var}_{P_0^i}(x_i)$, where $P_0^i$ is the prior associated with component $x_i$.
  Additionally, one initializes $w_\mu\InitIter=\augmeasure$. 
  Once the algorithm has converged, the minimum mean square error estimate of component
  $x_i$ is given by $a_i\ThisIter$. 
  %%% Guess not quite true if we include the noise update step
  %For $\damp=0$, cannonical AMP is recovered.
  }   
\label{algoCh1:AMP}
\end{figure}
Despite the fact that the augmented system matrix $\augsystem$ is sparse,
and that the derivation of AMP is based on dense matrices, nothing prevents us from applying it 
effectively to such a problem, and as we will see, it achieves state-of-the-art performance. 
The prior-dependent denoisers which iteratively produce the posterior estimates are given by
\begin{align}
  \fa(\Sigma_i^2, R_i; \omega) &\defas \myint{x} \!\! \!x\ \! P_0^i(x)\ \!\mathcal{N}\(x;R_i, \Sigma_i^2\) \notag\\
        &= R_i \ \! \mathbb{I}(i\le N) + \fasnipe(\Sigma_i^2, R_i;\omega) \ \! \mathbb{I}(i>N), \\
  \fc(\Sigma_i^2, R_i; \omega) &\defas \myint{x}\!\! \!
           x^2 \ \! P_0^i(x)\ \!\mathcal{N}\(x;R_i,\Sigma_i^2\) - \fa(\Sigma_i^2, R_i;\omega)^2 \notag\\
         &= \Sigma_i^2\ \! \mathbb{I}(i\le N) + \fcsnipe(\Sigma_i^2, R_i;\omega)\ \! \mathbb{I}(i>N)
\end{align}
where the prior terms are given by \eqref{eq:facPrior} 
and the SNIPE-specific denoisers \cite{BP2013} for the dual variables are given by
\begin{equation}
  \fasnipe(\Sigma^2, R;\omega) \defas \frac{R}{1 + e^{-\frac{R^2}{2\Sigma^2} + \omega}}, 
  \quad
  \fcsnipe(\Sigma^2, R;\omega) \defas \frac{1}{1 + e^{-\frac{R^2}{2\Sigma^2} + \omega}} \left( \frac{R^2}{1 + e^{\frac{R^2}{2\Sigma^2} - \omega}} + \Sigma^2 \right).
\end{equation}
\subsection{Learning the noise variance via the Bethe free energy}
\label{sec:learnnoise}
We now present the second key feature of \scampi, an iterative estimation of both the noise variance
and dual parameter relaxation, which improves reconstruction accuracy for natural images.
However, we note that this approach in fact \emph{degrades} reconstruction accuracy 
for strictly piecewise constant signals. 
Nevertheless, if the image of interest is \emph{known} to be 
strictly piecewise constant, the temperature $T$ can be fixed to $0$ to improve performance. If not, 
the hyperparameters $\{\Delta, T\}$ can be learned via the expectation-maximization we now detail.

Given the definition of the Bethe free energy $G$ derived in \cite{KMTZ14}
for the posterior measure \eqref{eq:pampd}, we update $\{\Delta, T\}$ at each iteration
according to their fixed-point values in $G$.
For our posterior, the Bethe free energy is written as
\begin{equation}
  G =\half\sum_{\mu}\[\frac{(\mathring{y}_\mu - \sum_i \mathring F_{\mu i} a_i)^2}{\mathring \Delta_\mu} + \log\(1+\frac{ \sum_i \mathring F_{\mu i}^2 v_i}{\mathring \Delta_\mu}\)+\log\(2\pi\mathring \Delta_\mu\)\] + \sum_i \dkl(P_i||P_0^i),
  \label{eq:BetheF_forAMP_last}
\end{equation}
where $\dkl(P_i||P_0)$ is the Kullback-Leibler divergence between the prior $P_0^i$ of the $i^{th}$ 
component and its estimated posterior 
$P_i(x_i|\Sigma_i^2,R_i) \propto P_0^i(x_i) \mathcal{N}(x_i;R_i,\Sigma_i^2)$, 
and the Gaussian fields $\bra{R_i, \Sigma_i^2}$ and posterior mean and variance 
estimates $\bra{a_i,v_i}$ of $x_i$ are the fixed-point 
values of the \scampi algorithm in Fig.~\ref{algoCh1:AMP}. 

For the moment, let us consider that there is 
a unique noise parameter $\augnoisevar$ for all of our factors. 
To minimize $G$, we start by isolating the terms which depend on $\augnoisevar$,
\begin{equation}
  G_{\augnoisevar} = 
    \frac{1}{2} \sum_{\mu=1}^{M+|E|}\[\frac{(\mathring{y}_\mu - \sum_{i=1}^{N+|E|} \mathring F_{\mu i} a_i)^2}{\augnoisevar} + \log\bigg(\augnoisevar+\sum_{i=1}^{N+|E|} \mathring F_{\mu i}^2 v_i\bigg)\] .
\end{equation}
We then write a fixed-point condition which is dependent on $\augnoisevar$, for all other 
variables fixed,
\begin{equation}
\frac{\partial G_{\mathring \Delta} }{\partial \mathring \Delta}
  = -\frac{1}{2}\sum_{\mu=1}^{M+|E|}\[\frac{1}{\mathring \Delta^2}\bigg(\mathring y_\mu - \sum_{i=1}^{N+|E|} \mathring F_{\mu i}a_i\bigg)^2 - \bigg(\mathring\Delta + \sum_{i=1}^{N+|E|} \mathring F_{\mu i}^2 v_i\bigg)^{-1}\] = 0.  \label{eq:toOptNoiseImages}
\end{equation}
It is possible to extract the optimal $\mathring\Delta$ by equating the two terms inside the sum, 
providing the set of solutions 
$\mathring \Delta_\mu \ \forall \ \mu \in\{1,\dots,M+|E|\}$. 
Defining the auxiliary term
$\chi_\mu \defas \mathring y_\mu - \sum_{i} \mathring F_{\mu i}a_i$,
each $\augnoisevaru$ is obtained by solving the quadratic equation
$\mathring \Delta_\mu^2 - \chi_\mu^2 \mathring \Delta_\mu -\chi_\mu^2\sum_{i}\mathring F_{\mu i}^2 v_i = 0$.
Since all variance terms are strictly positive, 
the feasible solution for this equation is simply $\augnoisevaru = g_\mu$ where
$g_\mu \defas \half\big(\chi_\mu^2 + \chi_\mu \sqrt{\chi_\mu^2+4\sum_{i}\mathring F_{\mu i}^2 v_i}\big)$.
Averaging over $\{\mathring\Delta_\mu:\mu\in\{1,\dots,M+|E|\}\}$, we obtain a single parameter 
$\mathring \Delta$. Recall, however, that we wish to consider two \emph{different} noise variances 
$\{\Delta,T\}$. Since both $\signal$ and $\dualvars$ exist in different domains, it is not 
reasonable to expect that they both should share the same noise variance estimate.
Instead, two averages are performed over the proper sets 
of variables in $\augsignal$.
Introducing the iterative time index, we obtain the fixed-point estimation
\begin{equation}
  \Delta\NextIter = \frac{1}{M}\sum_{\mu=1}^{M}g_\mu\ThisIter, 
  \quad
  T\NextIter = \frac{1}{|E|}\sum_{\mu=M+1}^{M+|E|}g_\mu\ThisIter, 
  \label{eq:noiseLearn}
\end{equation}
where $g_\mu\ThisIter$ is a function of $\bra{\Delta\ThisIter,T\ThisIter,\ba\NextIter,\bv\NextIter, \augmeasures, \augsystem}$. 
In the final implementation of \scampi, we also introduce a damping on the update of $\{\Delta,T\}$. 

Intuitively, this iterative estimation of the noise variances operates as a kind of 
annealing which plays a pivotal role in providing \scampi's superior reconstruction performance 
for natural images. Additionally, the estimation of $\{\Delta,T\}$ means that \scampi is more
robust to uncertainty on these hyperparameters than GrAMPA, 
an important feature for practical use cases. We present results demonstrating this effect 
in Sec. \ref{sec:results}. Our \scampi implementation can be downloaded on \url{https://github.com/jeanbarbier/scampi}.
\subsection{Hadamard operators for large-scale compressive imaging}
We now turn our attention to one important practical consideration for high-dimensional signals,
such as natural images. As $N$ increases, the construction, storage, and numerical use of
random, dense projection matrices containing $O(N^2)$ quickly becomes computationally infeasible.
To confront this issue, 
we use sub-sampled Hadamard operators, which have been empirically shown to provide 
performance very close to purely random matrices \cite{barbierSchulkeKrzakala}, 
but at a significant reduction in computational complexity, requiring 
$O(N\log N)$ operations for matrix multiplication.

Since each Hadamard mode, except 
the average mode, have exactly zero mean, the system \eqref{eq:hatSyst} would 
be invariant by a constant shift in the signal components without the presence
of the average mode in $\F$.
To break this symmetry, we force the retention of the average mode, which fixes 
the mean of the signal, when randomly selecting Hadamard modes for the 
construction of $\F$.
Once a realization of $\F$ is chosen (as well as $\bF^\intercal$ and $\bF^2, (\bF^2)^\intercal$ 
where the squared operators are simply sums in the case of Hadamard matrices), 
it is trivial to construct $\mathring{\tbf F}$ 
(and ${\mathring{\tbf F}}^\intercal, {\mathring{\tbf F}}^2, ({\mathring{\tbf F}}^2)^\intercal$, as 
well). Furthermore, $\mathring{\tbf F}$ does not generate memory issues since all sub-matrices,
other than $\F$, are extremely sparse with complexity of $O(N)$
per matrix multiplication. Thus, the overall complexity is $O(N\log N)$ per \scampi iteration.
\section{Experimental results}
\label{sec:results}
\begin{figure}[t]
  \centering
  \includegraphics[trim=1.5cm 0cm 2cm 0cm, clip=true, scale=0.22]{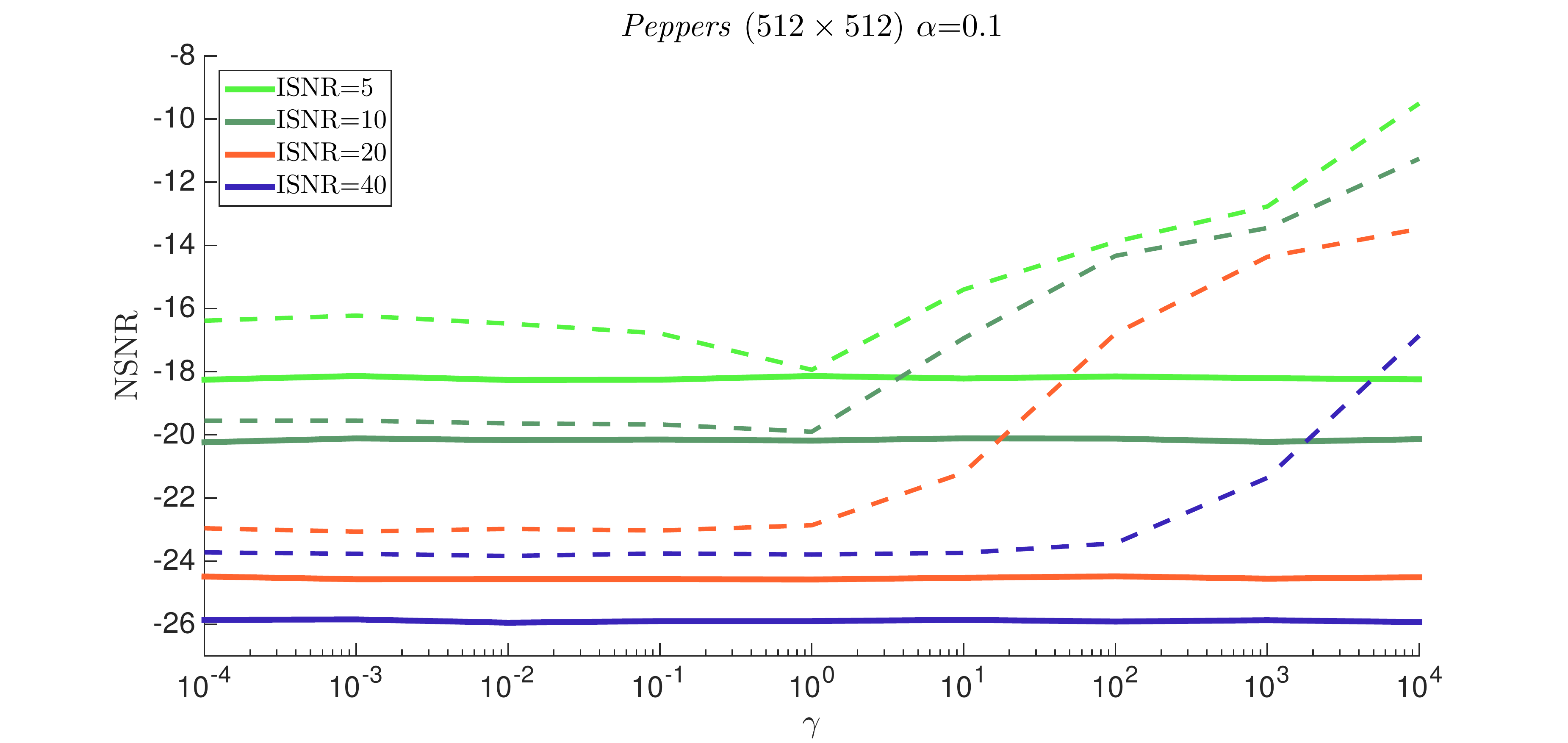}
  \includegraphics[trim=1.5cm 0cm 2cm 0cm, clip=true, scale=0.22]{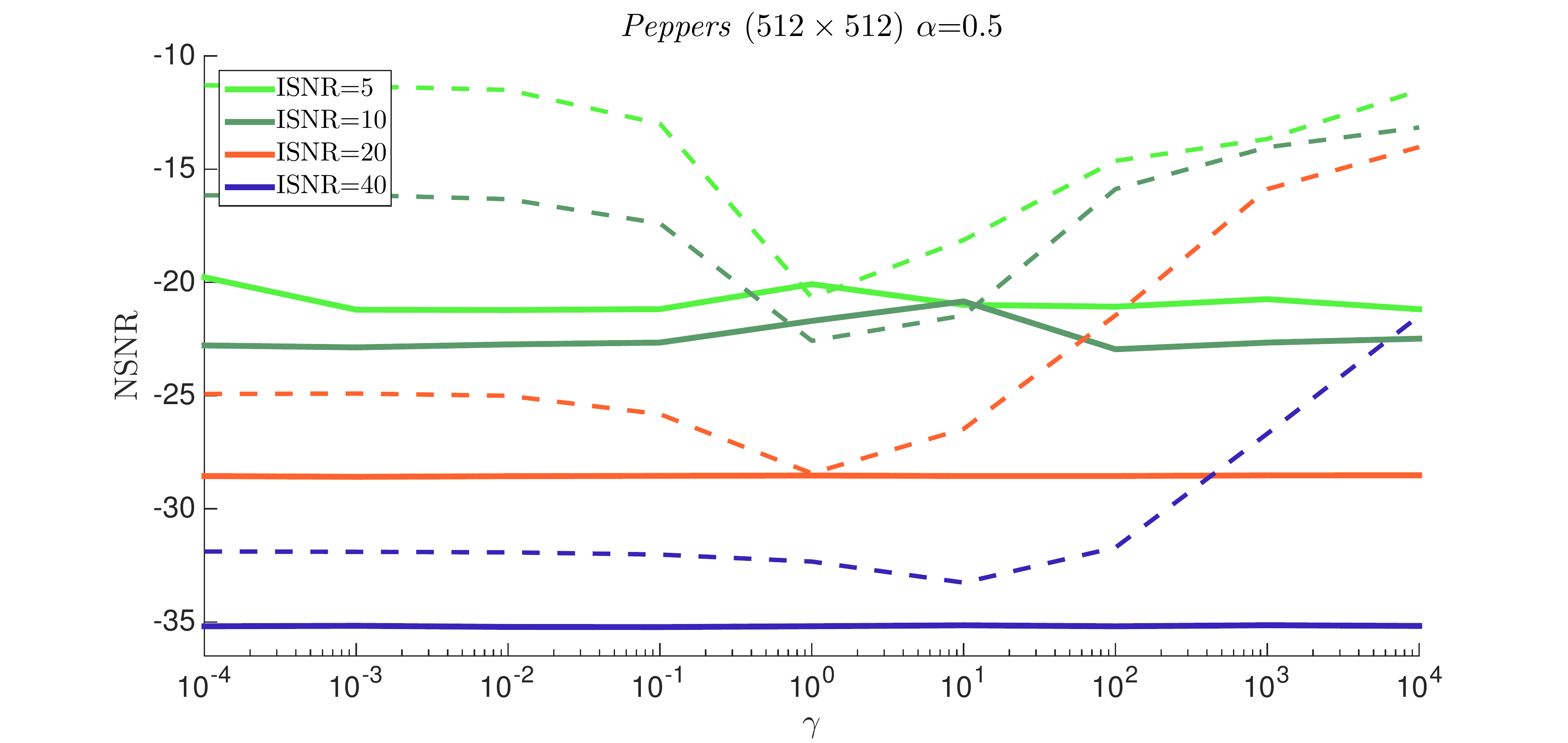}
  \includegraphics[trim=1.5cm 0cm 2cm 0cm, clip=true, scale=0.22]{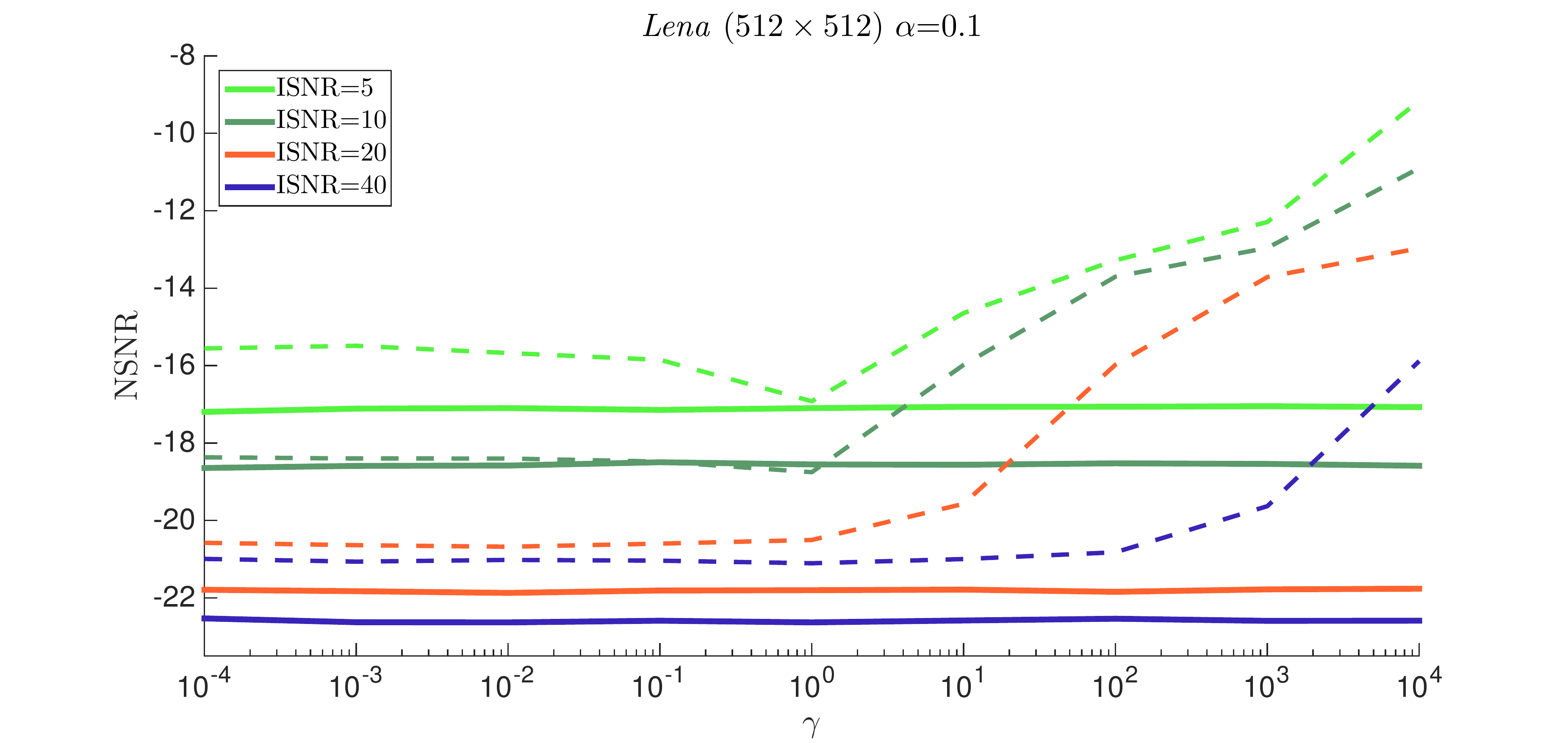}
  \includegraphics[trim=1.5cm 0cm 2cm 0cm, clip=true, scale=0.22]{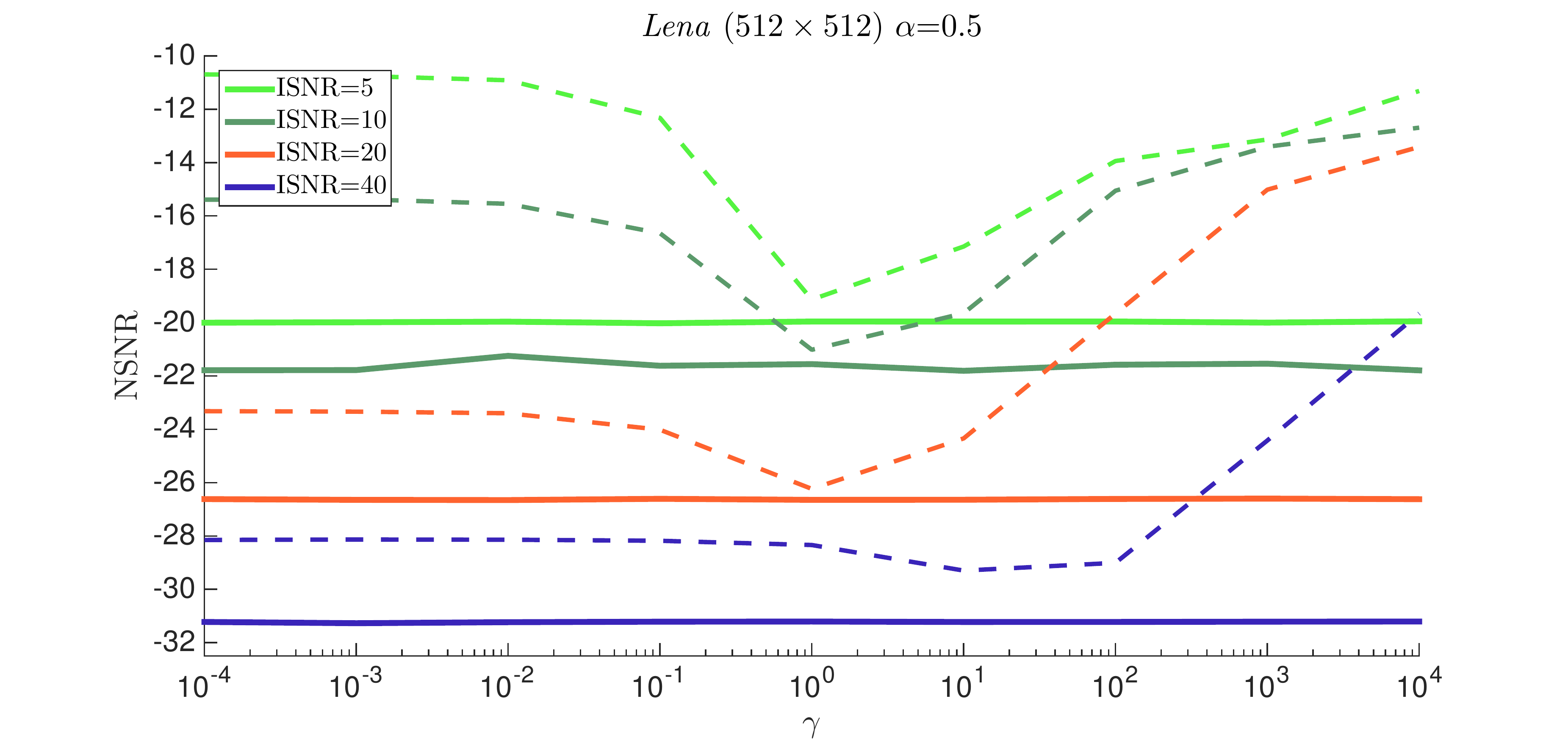}
  \includegraphics[trim=1.5cm 0cm 2cm 0cm, clip=true, scale=0.22]{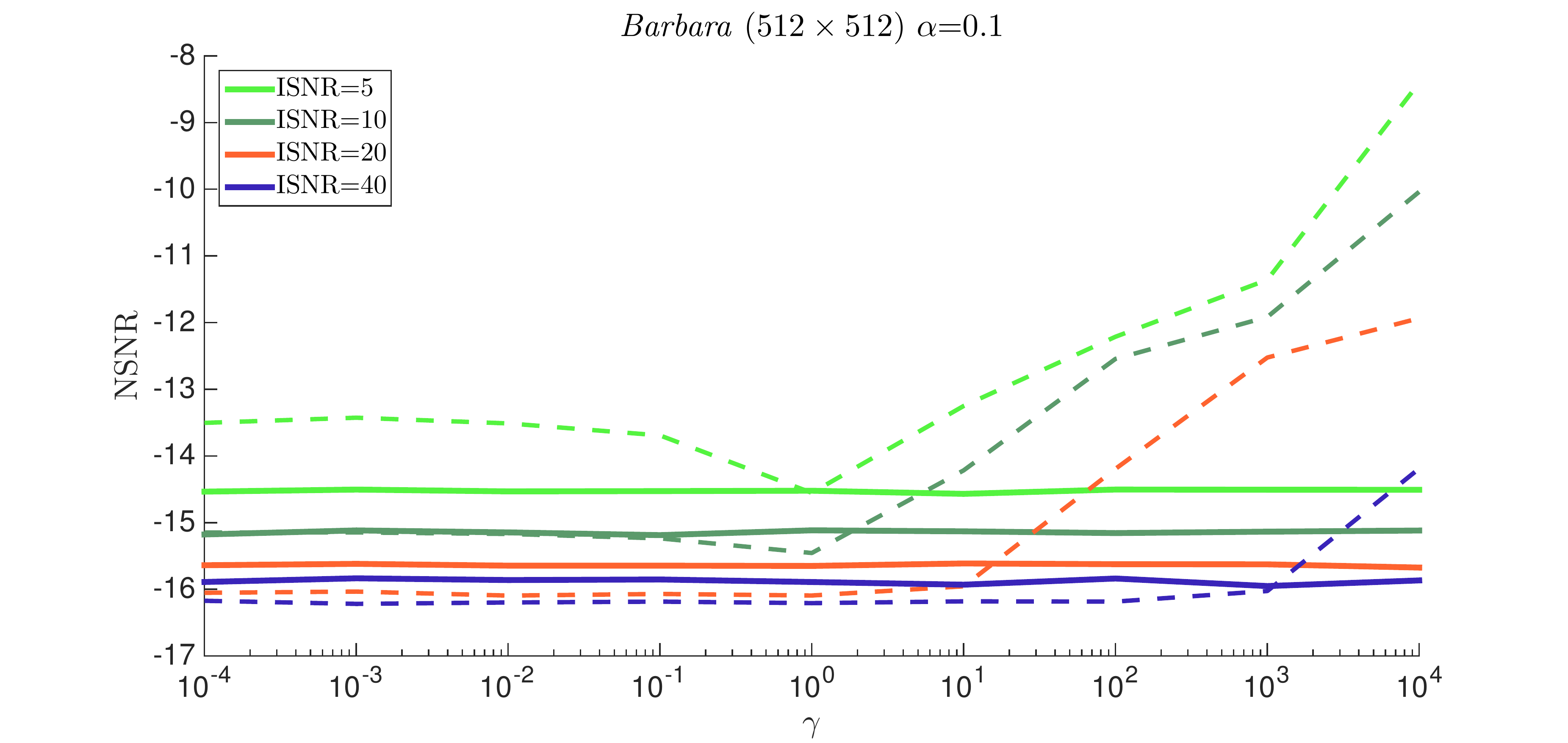}
  \includegraphics[trim=1.5cm 0cm 2cm 0cm, clip=true, scale=0.22]{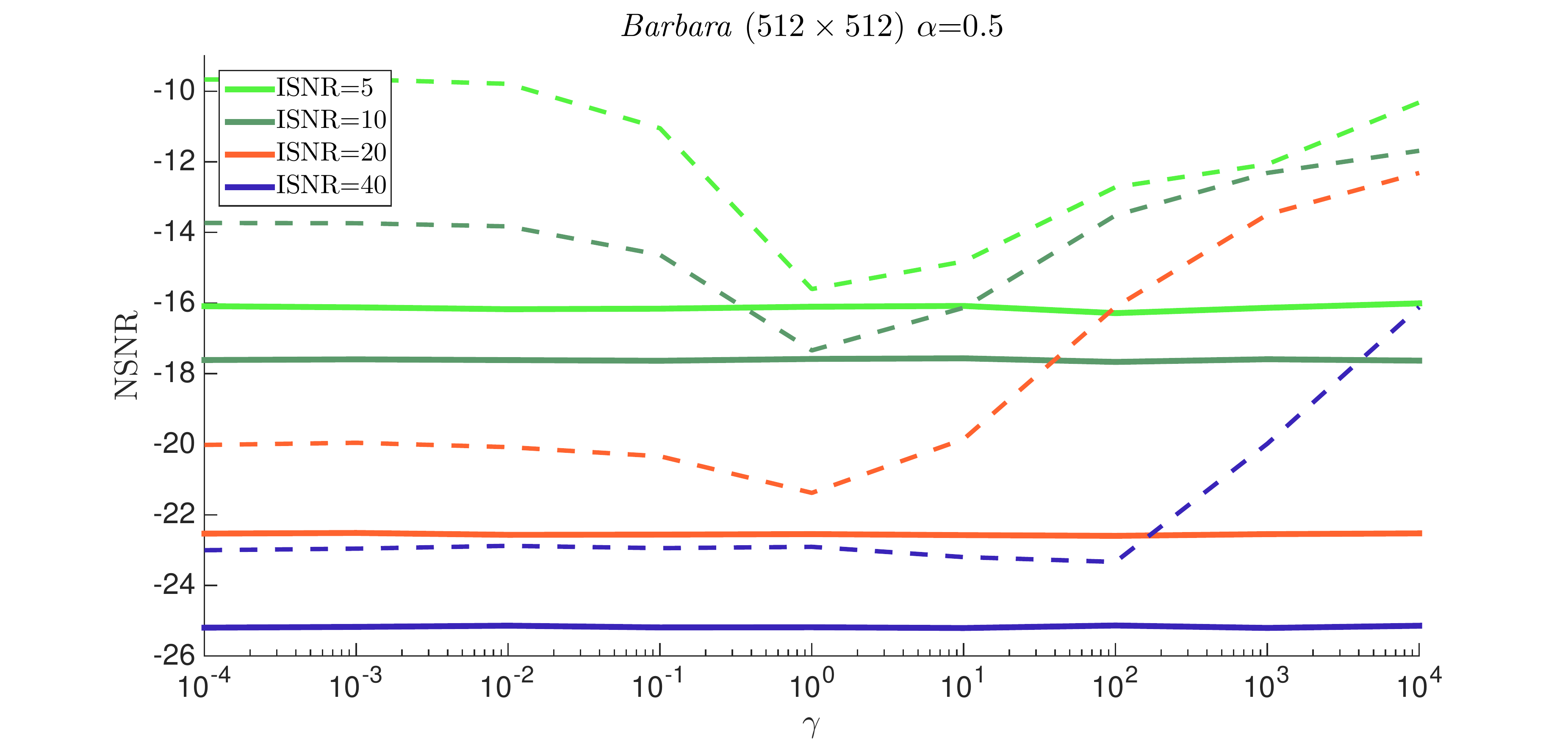}
  \includegraphics[trim=1.5cm 0cm 2cm 0cm, clip=true, scale=0.22]{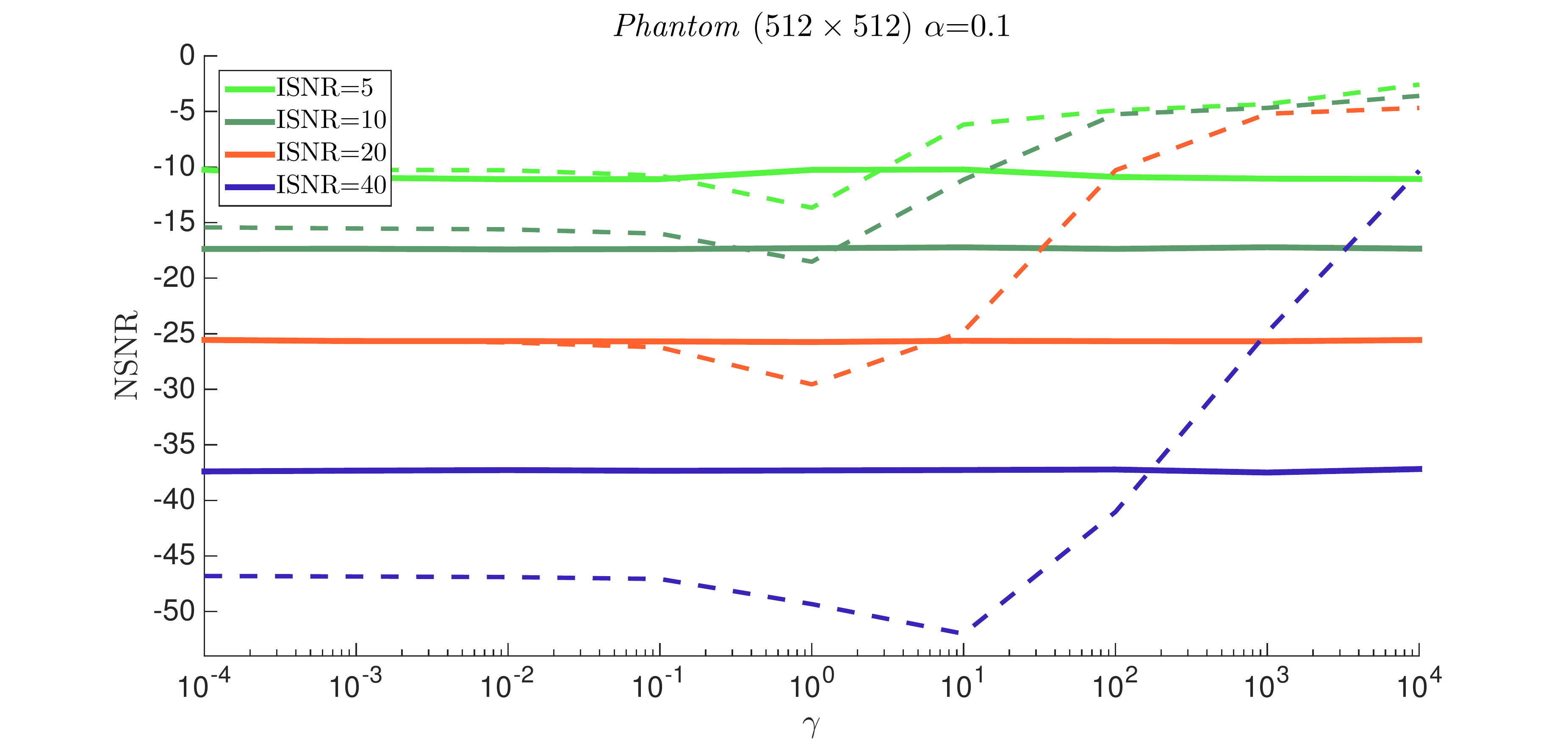}
  \includegraphics[trim=1.5cm 0cm 2cm 0cm, clip=true, scale=0.22]{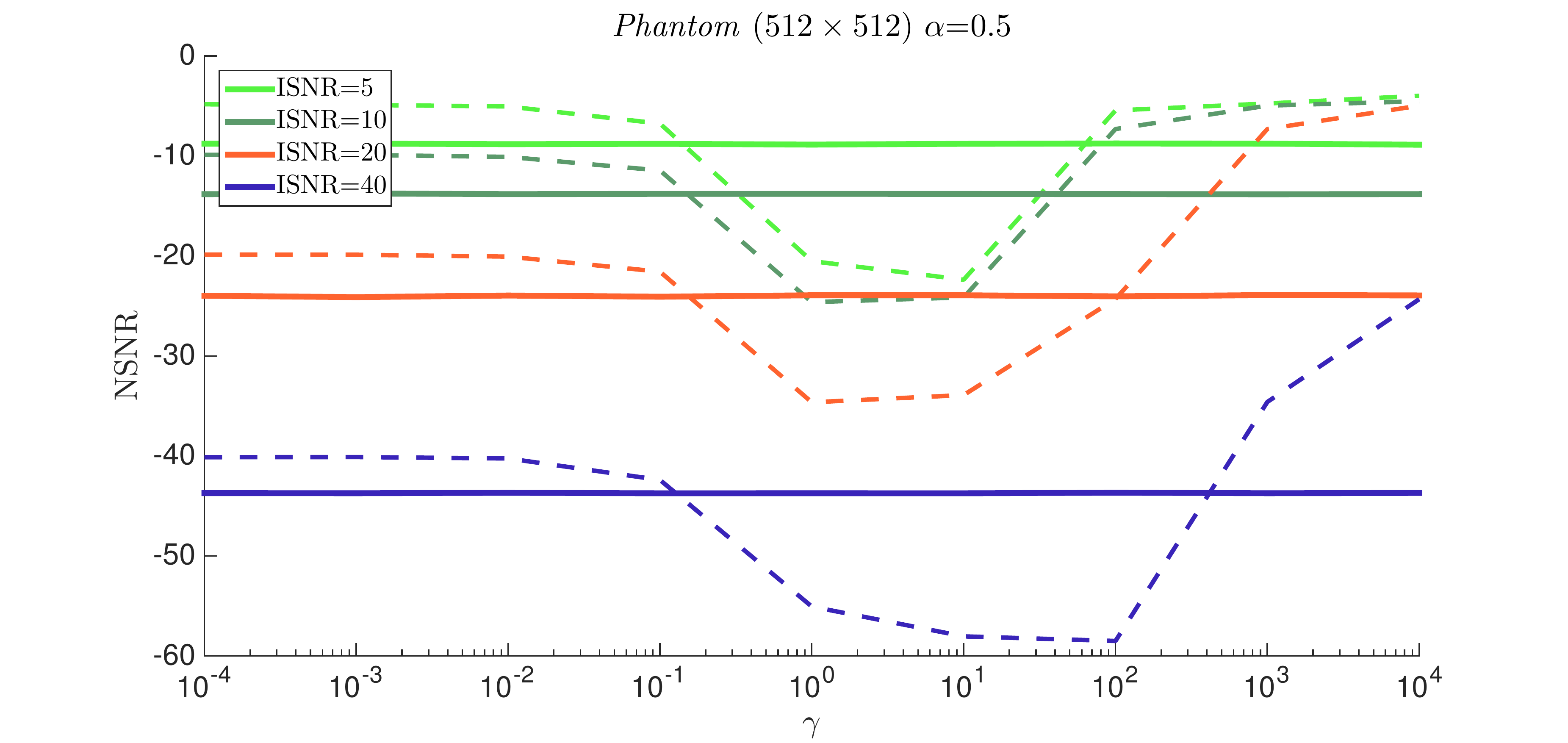}
  \caption{\label{fig:varRob}%
  Reconstruction performance in NSNR (the lower the better) of \scampi with noise learning 
  (solid curves) and GrAMPA (dashed curves), over the noise variance mismatch factor $\gamma$,
  for varying $512\times 512$ test images at $\alpha = 0.1$ (left) and $\alpha = 0.5$ (right).
  Colors correspond to various levels of ISNR (solid and dashed curves of
  same color correspond to same dB). Each point is obtained through a single run for a random 
  $(\textbf{F}, \bsy\xi)$ instance common to both algorithms. At this problem dimensionality,
  finite size performance fluctuations are negligible.}
\end{figure}
\begin{figure}[t]
  \centering
  \includegraphics[trim=7cm 0cm 7cm 0cm, clip=true, scale=0.43]{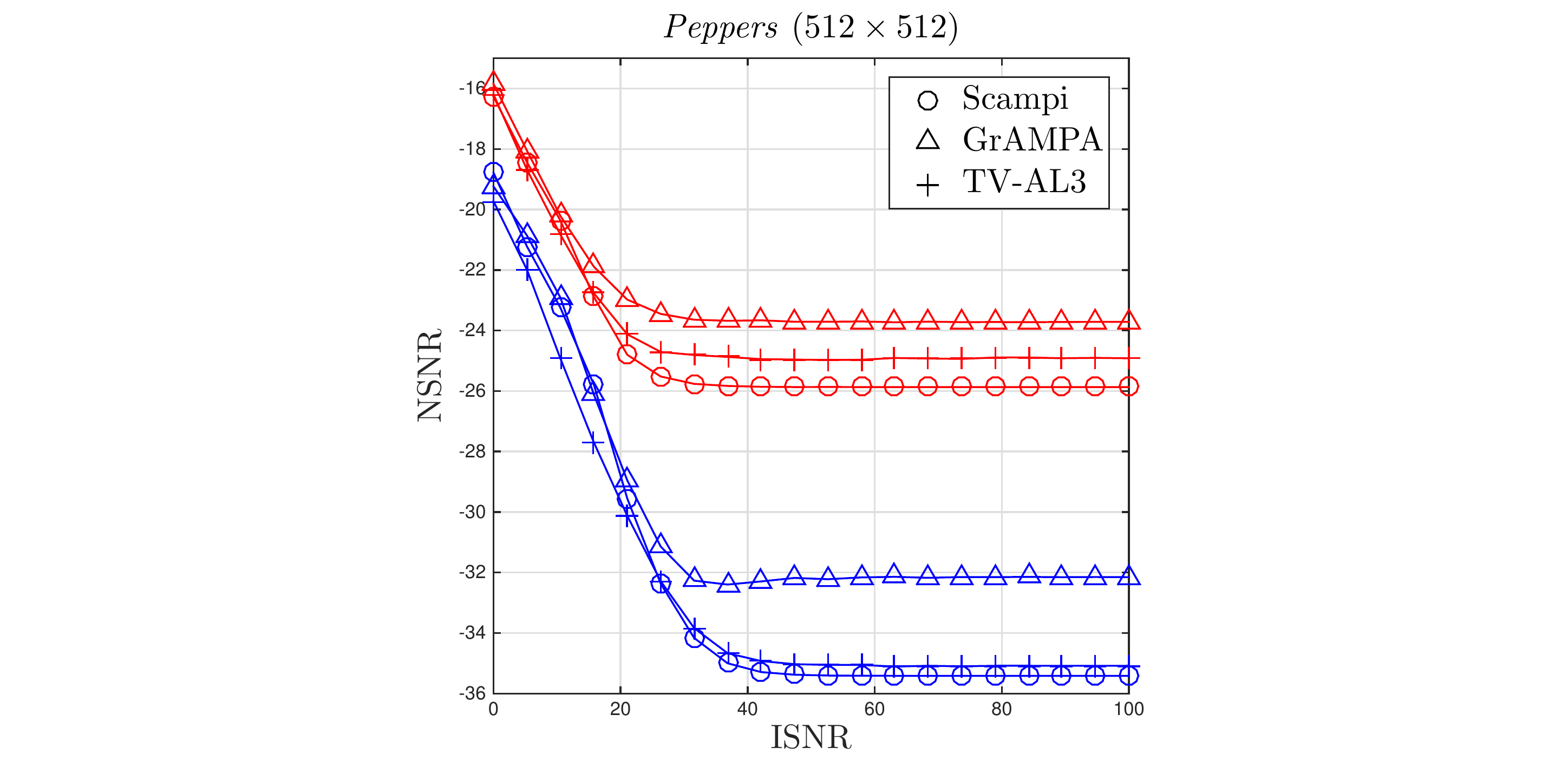}
  \includegraphics[trim=7cm 0cm 7cm 0cm, clip=true, scale=0.43]{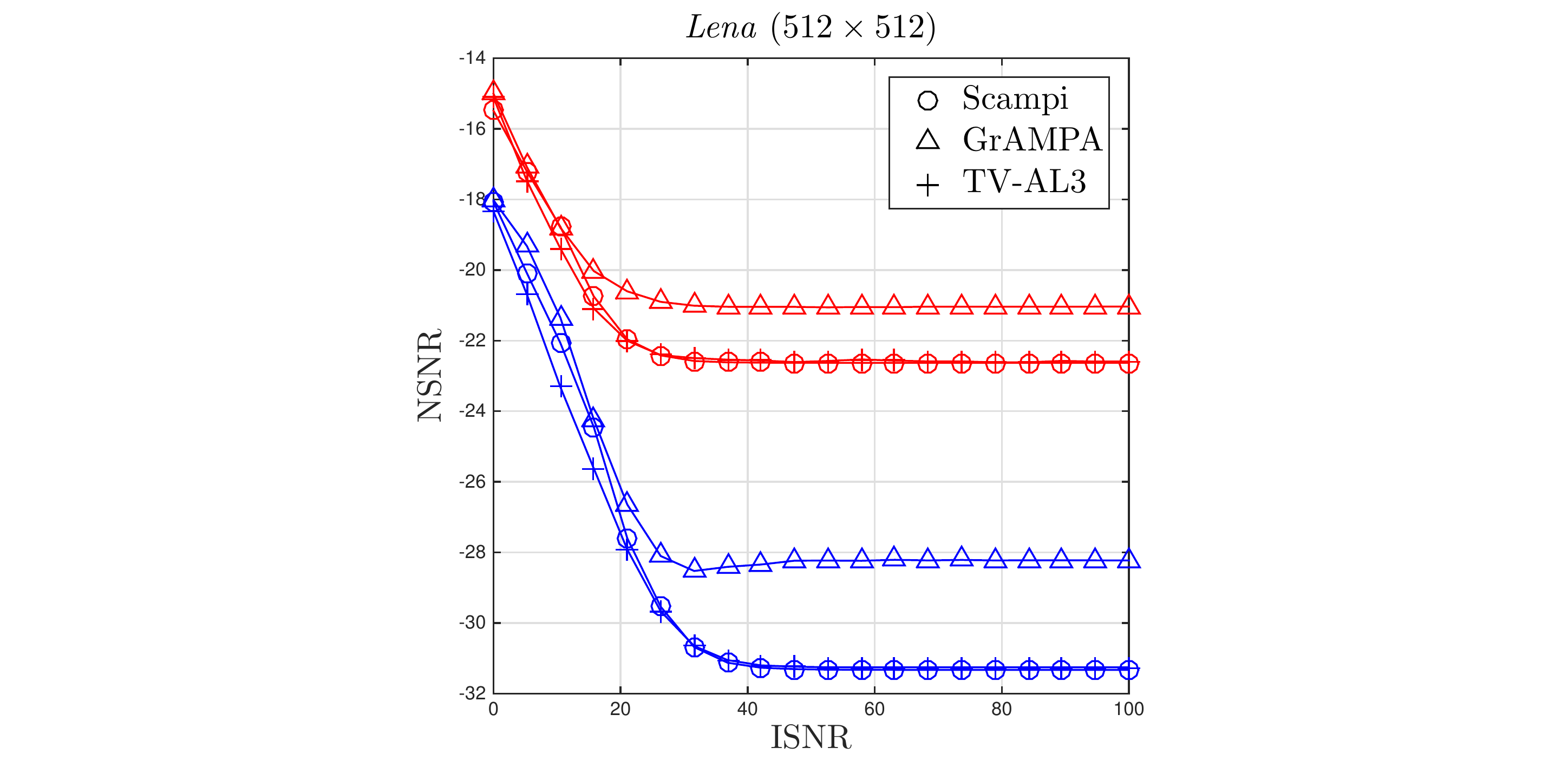}
  \includegraphics[trim=7cm 0cm 7cm 0cm, clip=true, scale=0.43]{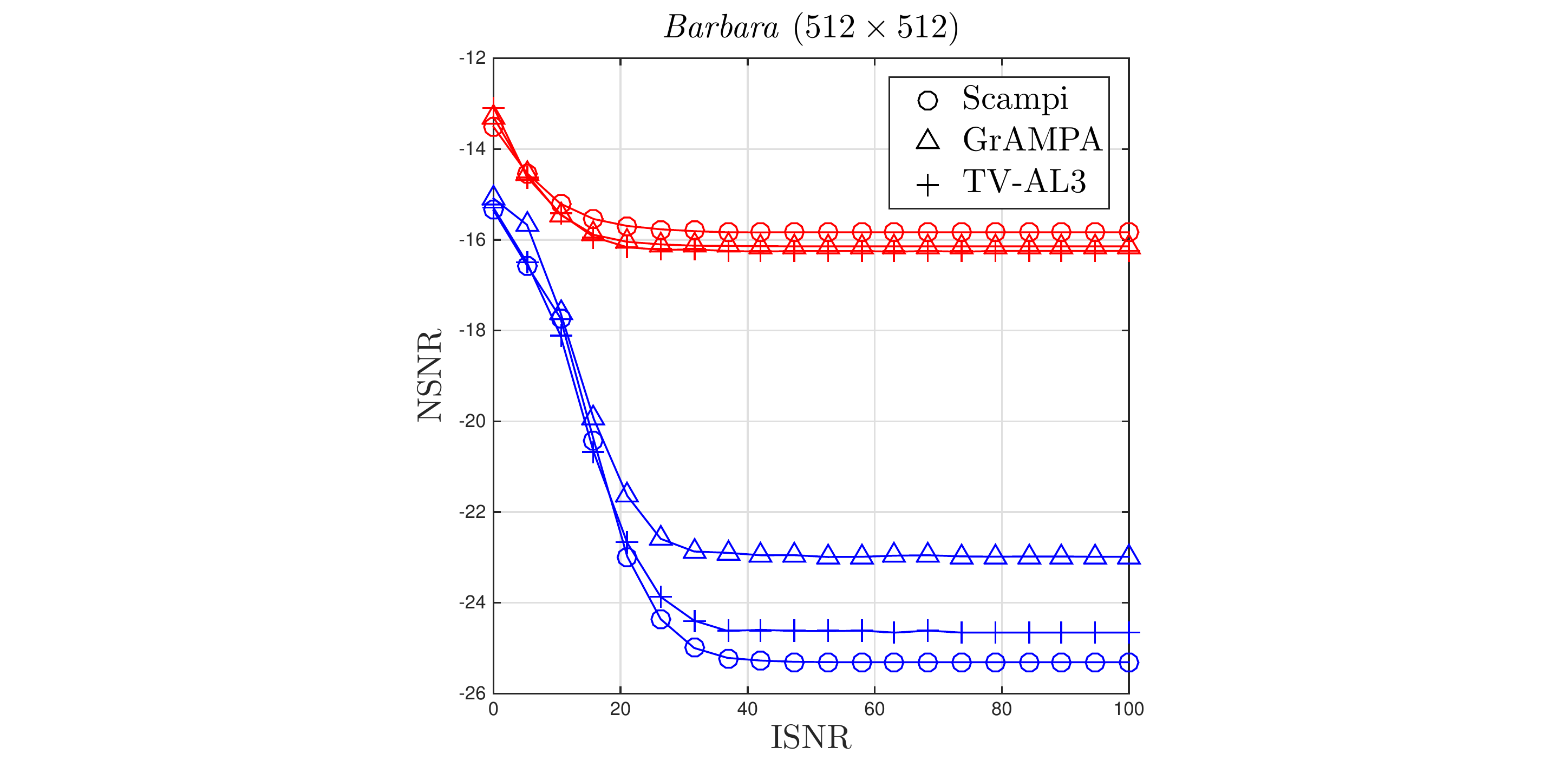}
  \includegraphics[trim=7cm 0cm 7cm 0cm, clip=true, scale=0.43]{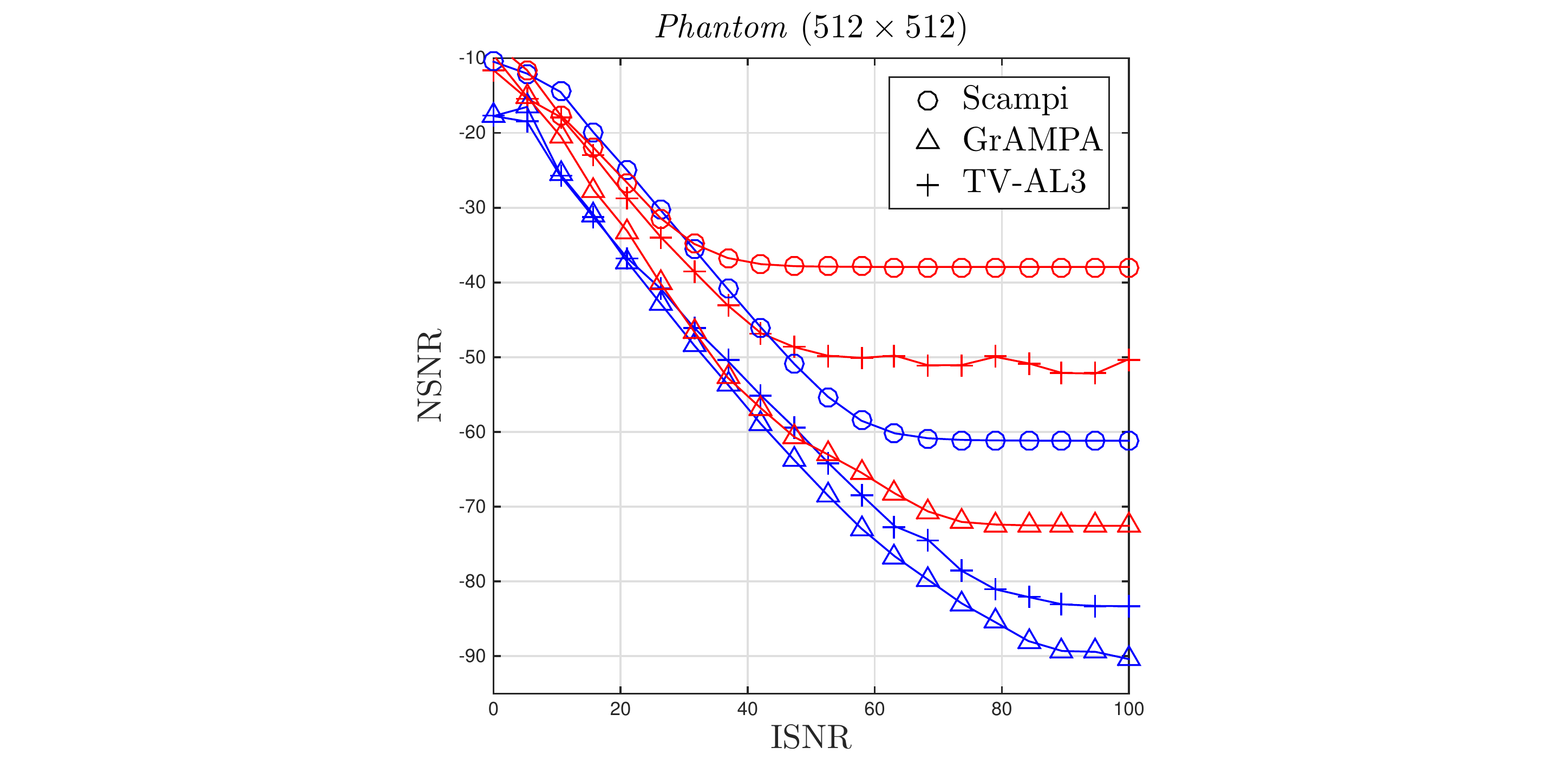}
  \caption{\label{fig:optimal_snr}%
     Comparison of optimal reconstruction accuracy
     as a function of the ISNR for Scampi, GrAMPA, and TV-AL3. 
     All test points correspond to a search over 30 test-points for tuning the regularization
     parameter of TV-AL3 or the SNIPE $\omega$ parameter for GrAMPA and Scampi. Two measurement
     rates are tested for each method on each $512\times 512$ image, $\alpha = 0.5$ (blue) and $\alpha = 0.1$ (red).}
\end{figure}
We now present the results of two experimental protocols for image reconstruction from 
AWGN corrupted CS measurements. In all cases, the tested images are
rescaled to the range $s_i\in [0,1]$, though this property isn't leveraged by any of the tested
algorithms. Reconstruction performance is quantified 
using the normalized signal to noise ratio (NSNR), measured in decibels, between the 
true image $\plantedsignal$ and the reconstruction $\hat{\signal}$ defined as 
$\text{NSNR}(\hat{\signal},\plantedsignal)\defas 10 \log_{10}(||\plantedsignal - \hat{\signal}||_2^2 / ||\plantedsignal||_2^2)~{\rm dB}$. 
For all experiments, the level of the true experimental noise variance, $\Delta^*$, 
is reported as the \emph{input} SNR (ISNR) and calcualted as
$\text{ISNR}=10\log_{10}(||\textbf{y}||^2_2/(M\Delta^*))~{\rm dB}$.

We conduct our experiments for the common test images \emph{Lena}, \emph{Barbara},
\emph{Peppers}, as well as the Shepp-Logan phantom (\emph{Phantom}) at a resolution of 
$512\times 512$ pixels. \emph{Barbara} and
\emph{Lena} represent common examples of complex natural images, with \emph{Lena} possessing a
mixture of both
smooth and high-variation regions, while \emph{Barbara} contains very many edges due to the 
large amount of textured content in the image. \emph{Peppers} represents a simpler case of a
more piecewise-continuous natural image, and \emph{Phantom} is, simply, a cartoon image of 
a few regions of \emph{constant} pixel value.

The first experiment, which results are presented in Fig.~\ref{fig:varRob}, compares the robustness of GrAMPA 
and \scampi, which includes the noise learning (\ref{eq:noiseLearn}), to uncertainty of the measurement AWGN noise variance $\Delta^*$. 
To simulate this uncertainty, both GrAMPA and \scampi are provided with an initial
estimate of the noise variance, $\Delta = \gamma\Delta^*$, where $\gamma=1$ represents
perfect knowledge about the noisy channel.
Each plot corresponds to the achieved reconstruction performance for a different image and two
sub-sampling rates $\alpha = M/N \in \{0.1, 0.5\}$, as a function of the mismatch factor $\gamma \in [10^{-4}, 10^{4}]$. The dashed and solid curves correspond to the best GrAMPA and \scampi performances, respectively,
over the range of the parameter
$\omega\in\{0,1,\dots,5\}$ for the SNIPE prior $\snipeprior(\cdot;\omega)$. 
Generally, the best $\omega$ value for GrAMPA is the same (or close) to the one for \scampi. 
As seen in these charts, \scampi is able to provide better NSNR performance than GrAMPA, and
without requiring any specific knowledge about $\Delta^*$. GrAMPA, however, requires that 
$\Delta^*$ be known, in most cases, within an order of magnitude. Additionally, in the case of
images, such as \emph{Phantom}, for high ISNR, 
exact knowledge of $\Delta^*$ is in fact \emph{detrimental} to the performance of the algorithm.

In the second experiment, see Fig.~\ref{fig:optimal_snr}, 
we show the best-case performance of both GrAMPA and \scampi 
as a function of the ISNR for the specified test images at $\alpha = 0.1$ and $\alpha = 0.5$. 
We compare both approaches to TV-AL3 \cite{Li2009}, as well, to give a reference point between
these two probabilistic techniques and a state-of-the-art convex technique. For these experiments,
best-case means that we intensively test over the free parameters of the algorithms, reporting only
the best results. In the case of GrAMPA and \scampi, we test over $\omega$, and for TV-AL3 we test
over the weight of the TV regularization cost. In all cases, each point on the curves represents 
the best outcome over 30 reconstructions sweeping over each tuning parameter's domain. 
Additionally, we provide all algorithms with the true $\Delta^*$.
Our goal in reporting these best-case results is to compare the \emph{maximal} performance of 
these techniques.

The findings we report are quite interesting and varied. The first topic we address is that 
for all tested natural images, TV-AL3 performs better than GrAMPA in terms of NSNR. Only
for \emph{Phantom}, whose gradient is very well modeled by the SNIPE prior, does GrAMPA 
outperform TV-AL3. However, one should note that the tuning of the free parameters of TV-AL3
is more problematic \emph{a posteriori} than GrAMPA, as the domain of the optimal regularization
term can be quite large, depending on the problem instance, while the optimal value of $\omega$
is generally around $0$ or $1$. But, these maximally best-case 
performance results imply that the convex treatment of TV is superior to the probabilistic model
used by GrAMPA in the case of natural images.

However, the model used by \scampi, along with the dynamic estimation of both the measurement
noise variance and constraint slack on the dual variables, provides best-case performance equal to or beyond TV-AL3's ones for most test cases, notably on \emph{Peppers} and \emph{Lena}, and at high 
measurement rates for an edge-heavy image such as \emph{Barbara}. Finally, as we pointed out
earlier, we see the expected subpar performance of \scampi for the cartoon \emph{Phantom} image.
However, we did not test the fixed $T=0$ case for this image, which may yet improve the performance
of \scampi for such images. 

\begin{figure}[t]
  \centering  
  \includegraphics[trim=12cm 1cm 12cm 1cm, clip=true, scale=0.2]{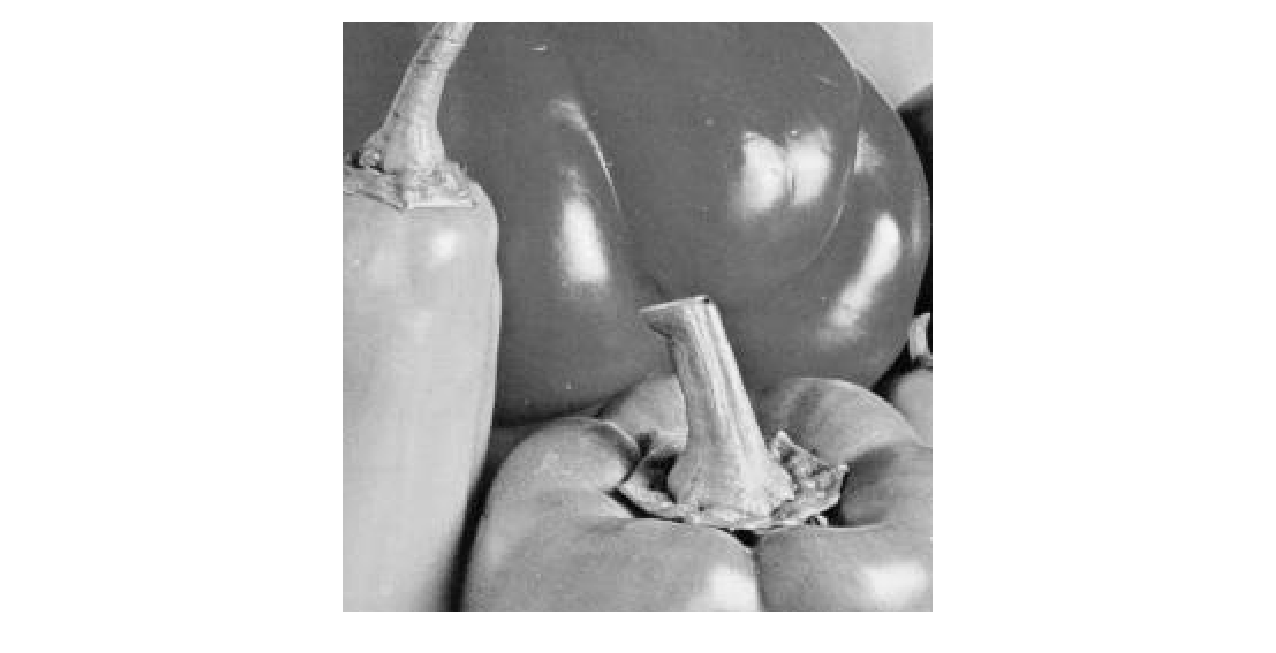}
  \includegraphics[trim=12cm 1cm 12cm 1cm, clip=true, scale=0.2]{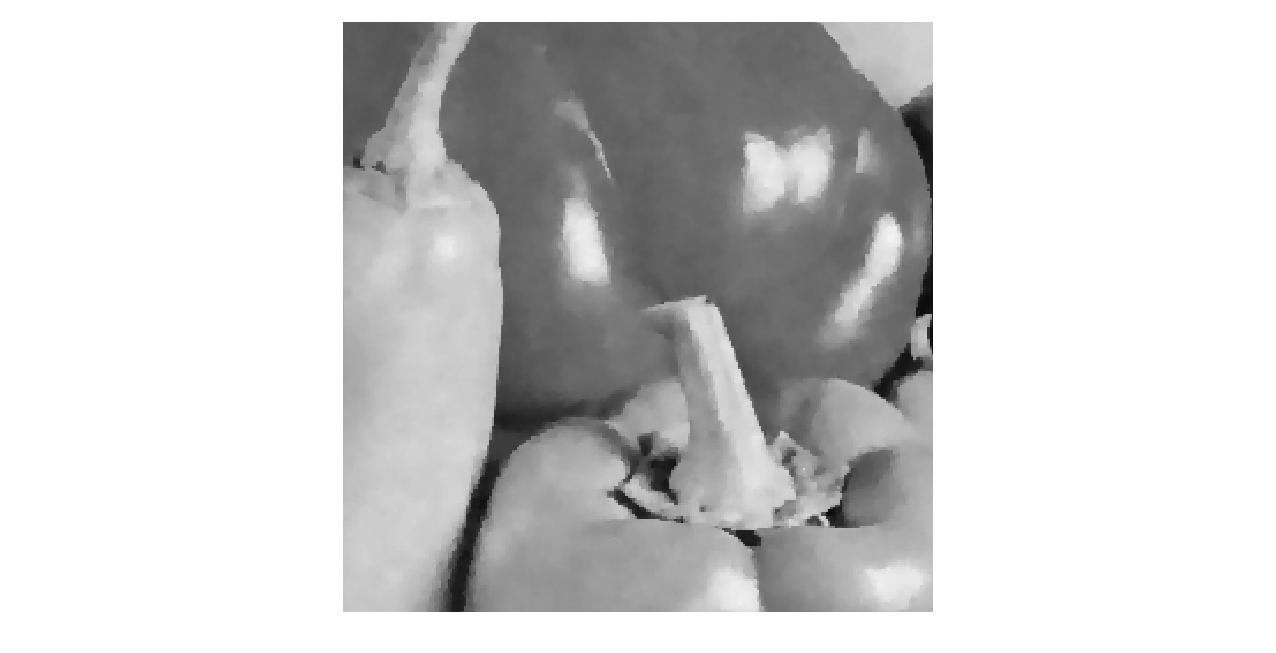}
  \includegraphics[trim=12cm 1cm 12cm 1cm, clip=true, scale=0.2]{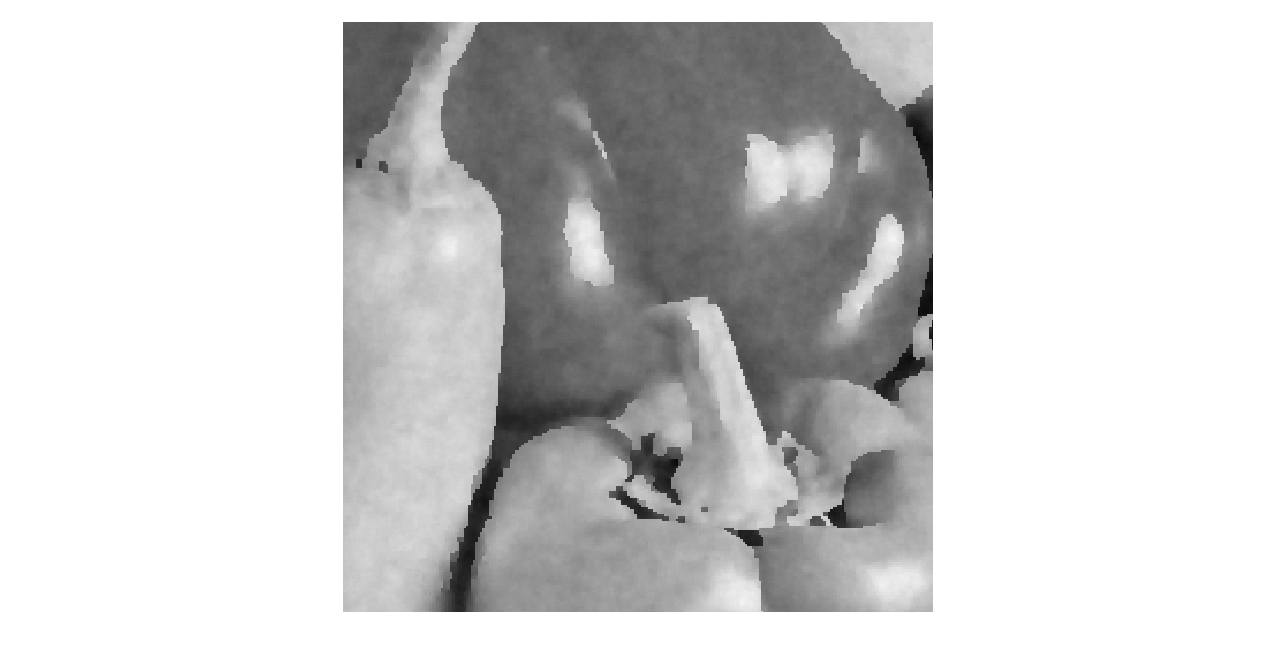} 
  \includegraphics[trim=12cm 1cm 12cm 1cm, clip=true, scale=0.2]{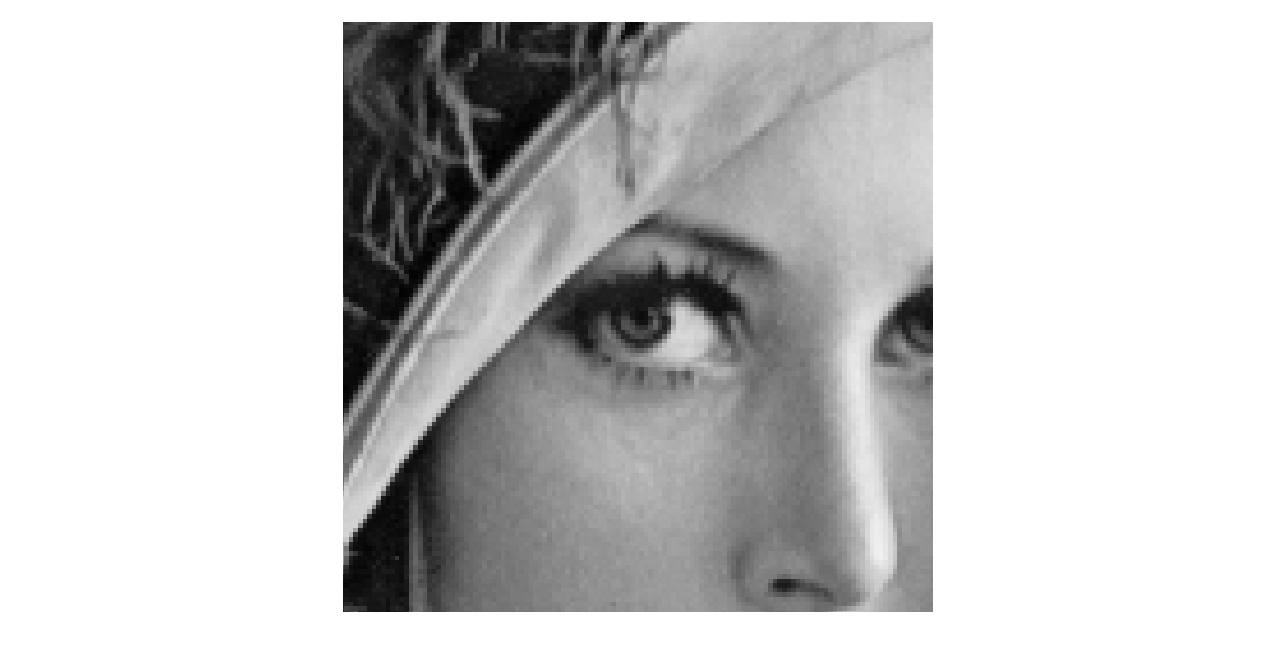}
  \includegraphics[trim=12cm 1cm 12cm 1cm, clip=true, scale=0.2]{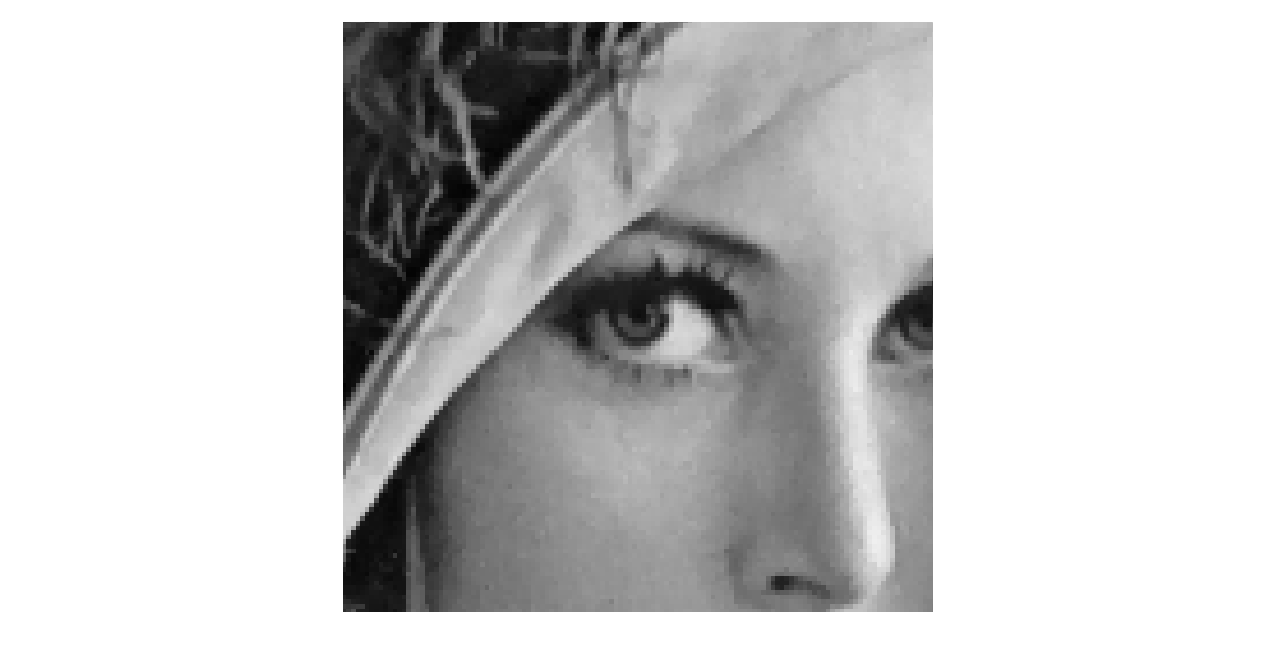}
  \includegraphics[trim=12cm 1cm 12cm 1cm, clip=true, scale=0.2]{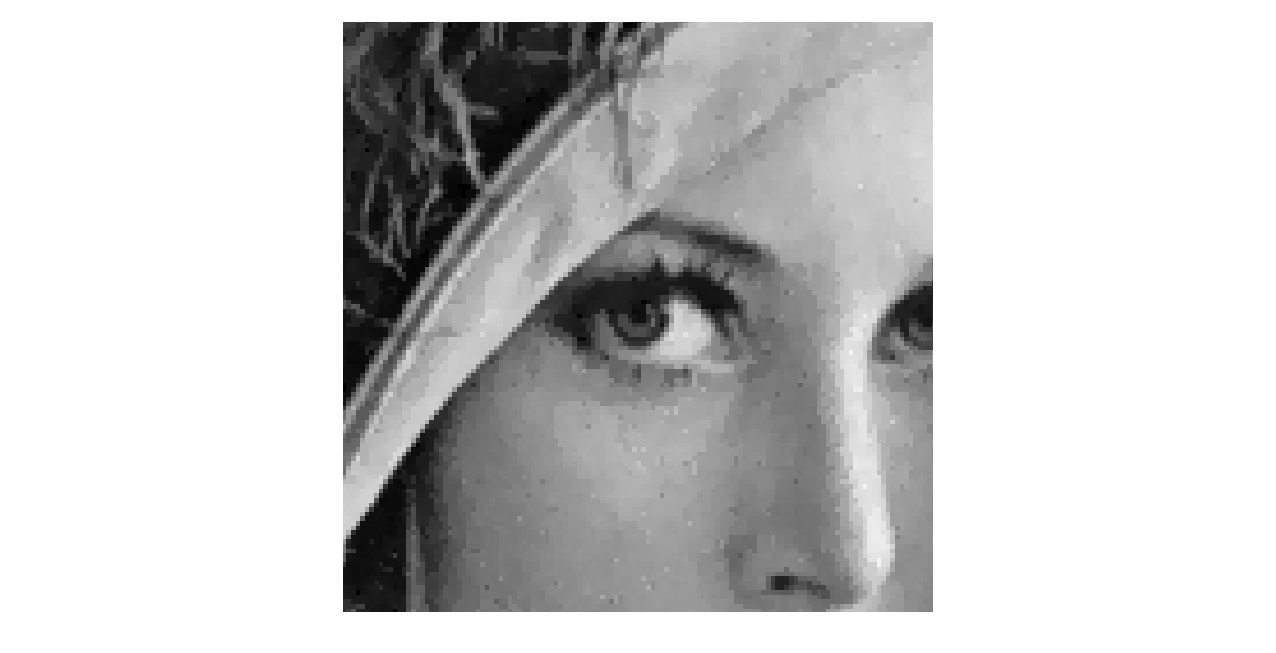}   
  \includegraphics[trim=12cm 1cm 12cm 1cm, clip=true, scale=0.2]{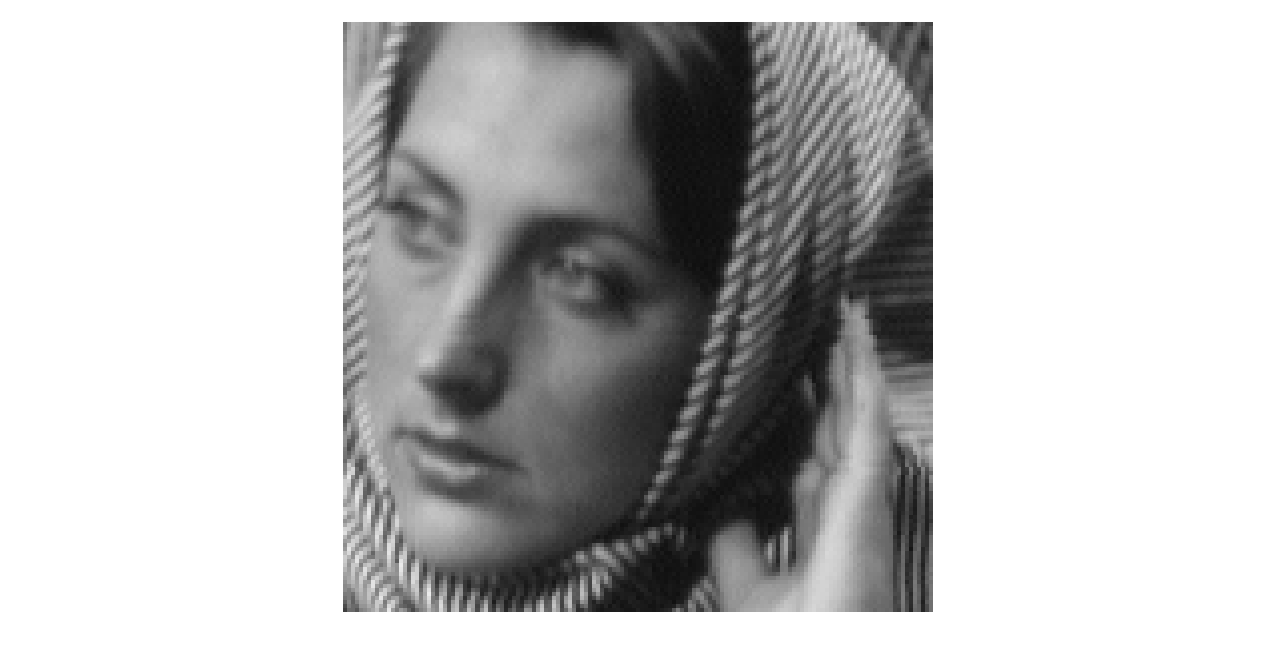}
  \includegraphics[trim=12cm 1cm 12cm 1cm, clip=true, scale=0.2]{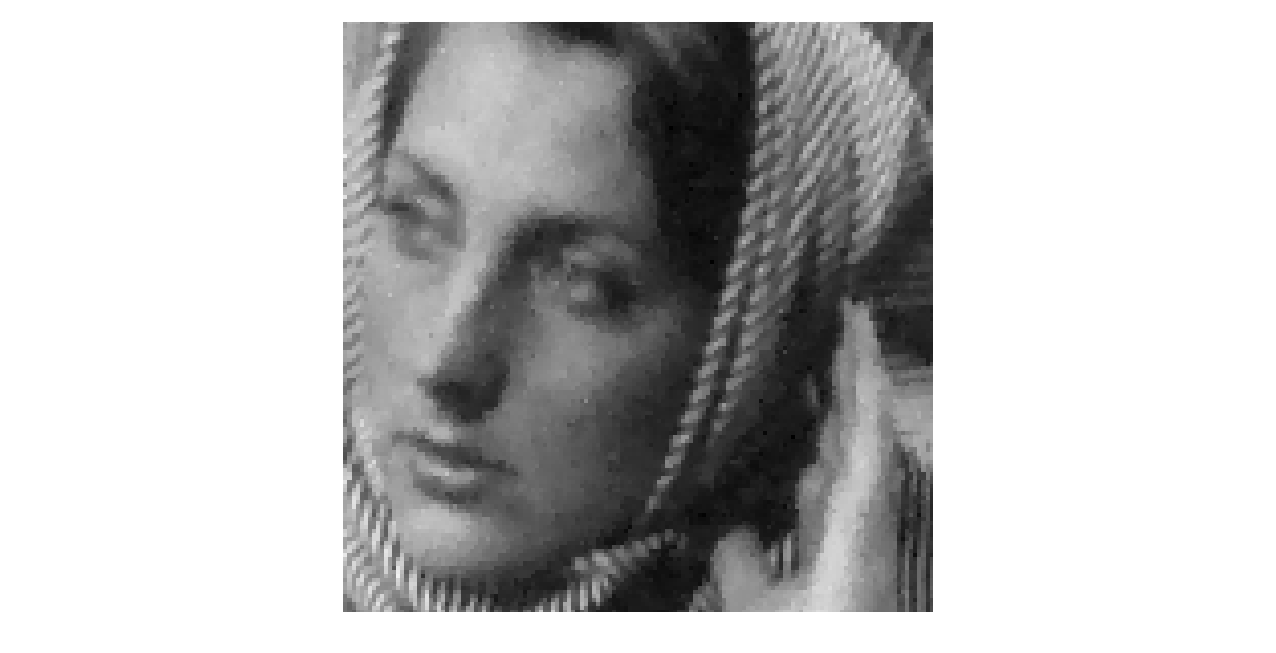}
  \includegraphics[trim=12cm 1cm 12cm 1cm, clip=true, scale=0.2]{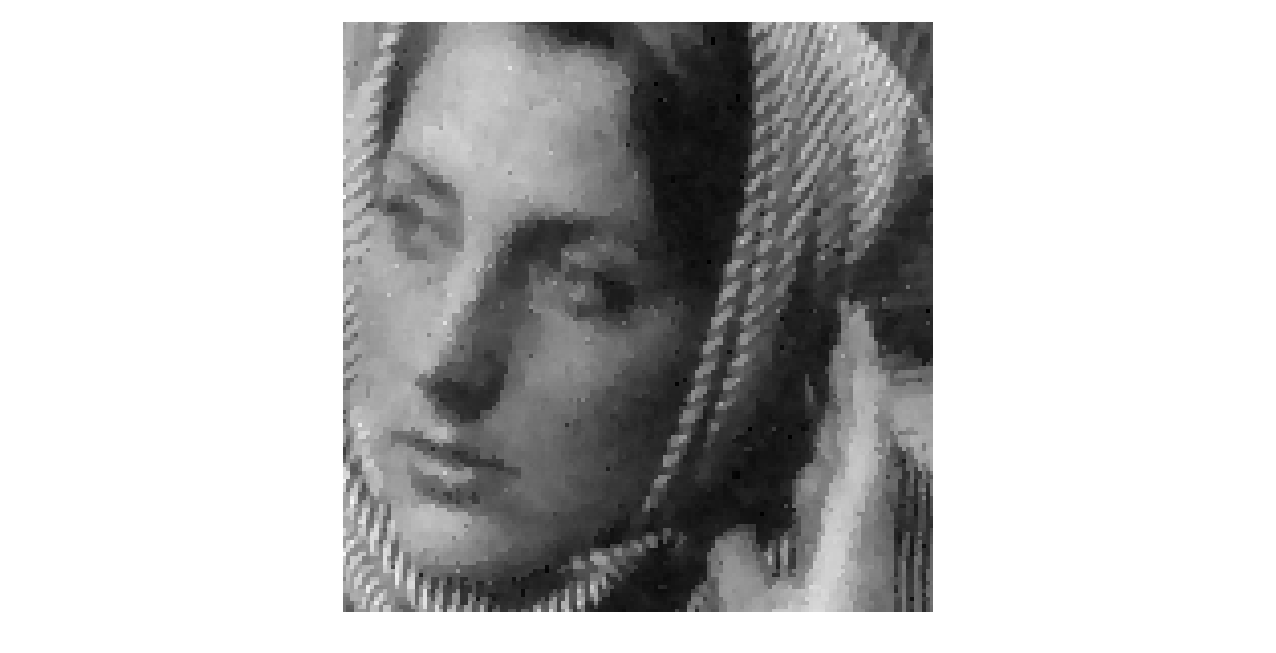} 
  \includegraphics[trim=12cm 1cm 12cm 1cm, clip=true, scale=0.2]{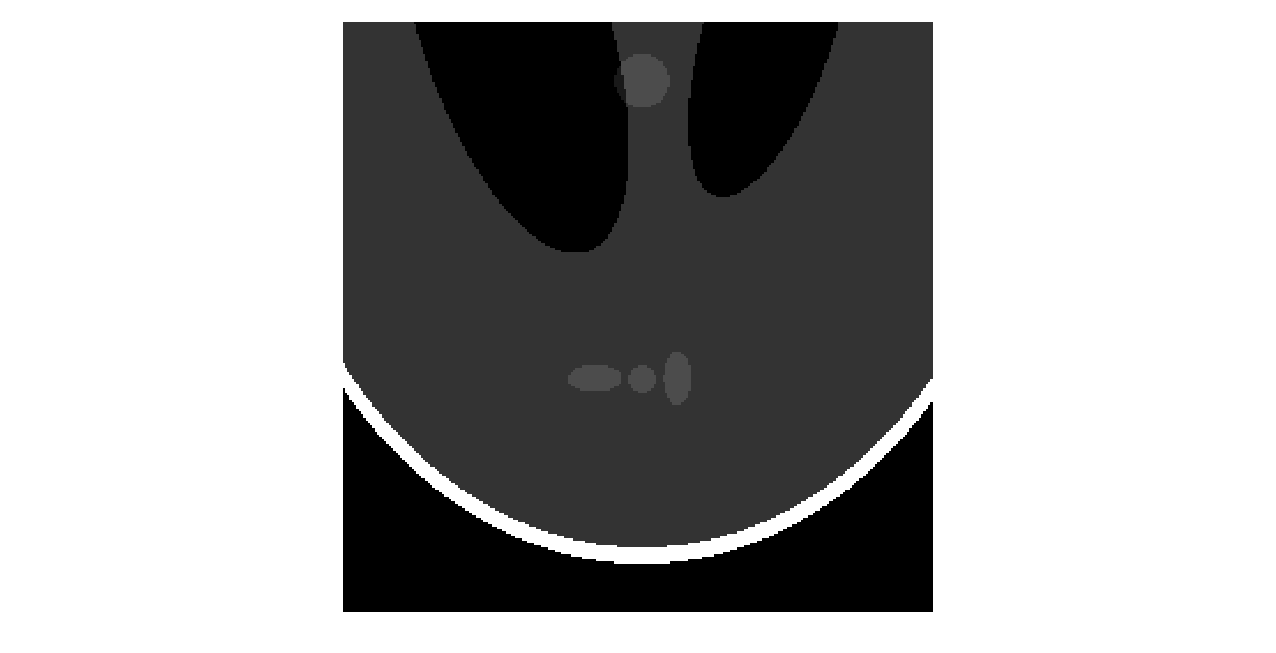}
  \includegraphics[trim=12cm 1cm 12cm 1cm, clip=true, scale=0.2]{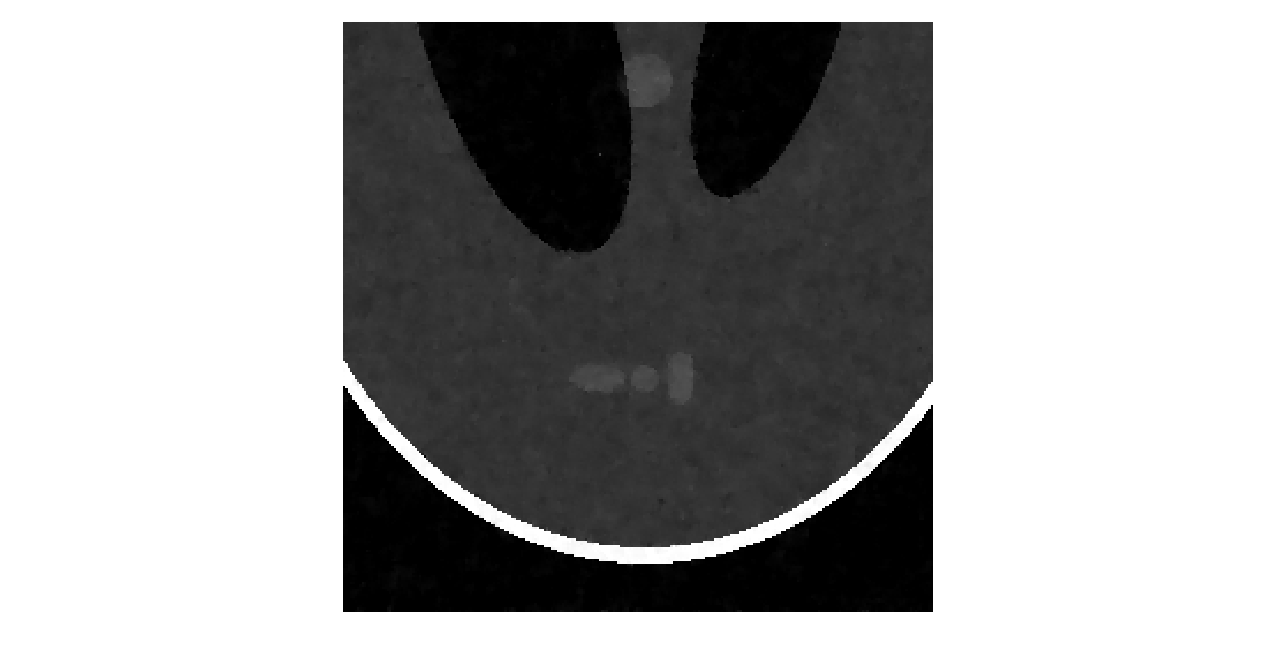}
  \includegraphics[trim=12cm 1cm 12cm 1cm, clip=true, scale=0.2]{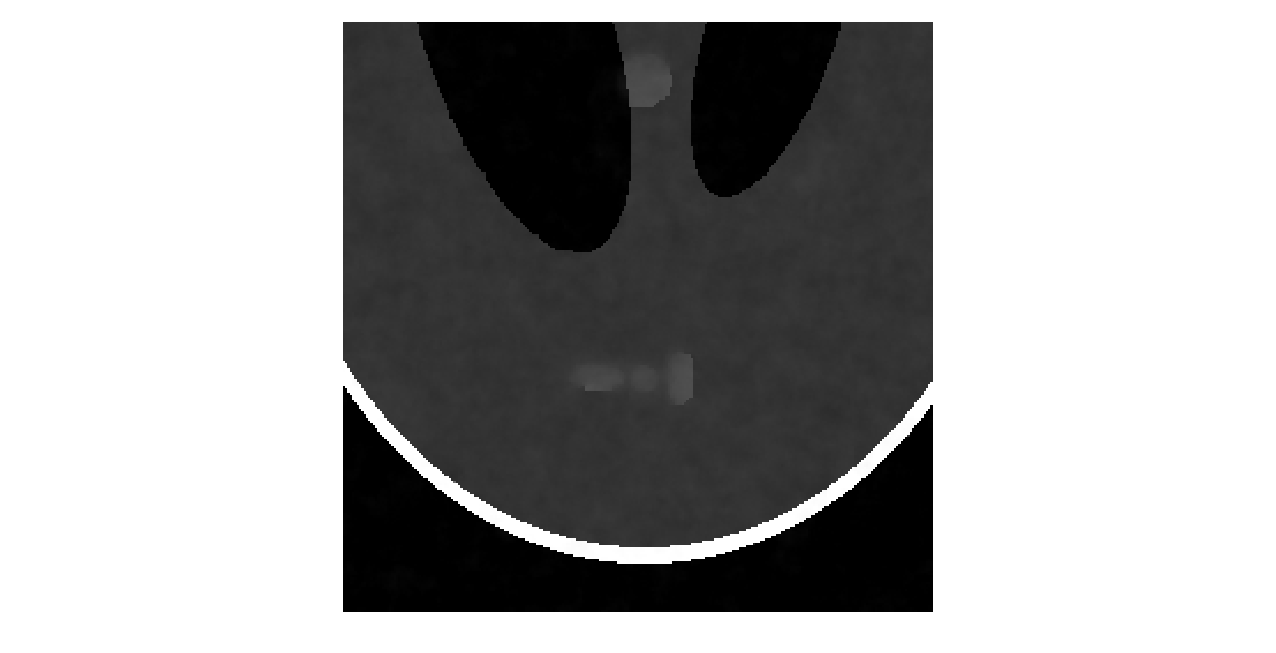} 
  \caption{\label{fig:visual_compare}%
     Visual comparison for some $512\times 512$ images: a zoom on the original image (left) is compared with the Scampi (center) and GrAMPA (right) 
     reconstructions for different settings, all with a fixed SNIPE prior parameter $\omega=0$. From top to bottom are \emph{peppers} at $\alpha = 0.1$ for an ISNR of 20~dB, \emph{Lena} at $\alpha = 0.5$ for an ISNR of 40~dB, \emph{Barbara} at $\alpha = 0.5$ for an ISNR of 30~dB and \emph{phantom} at $\alpha = 0.1$ for an ISNR of 20~dB.}
\end{figure}

We also show on Fig.~\ref{fig:visual_compare} some reconstruction results obtained with \scampi and GrAMPA for different settings, without optimization of $\omega$ which is set to $0$ in all cases. As expected, the images obtained with \scampi are smoother and match better the three first natural images. In the last case of \emph{phantom}, the gain in NSNR with GrAMPA with respect to \scampi is due to the large constant areas, but we notice that even in this case, \scampi gets a better resolution at the edges.
\section{Conclusion}
\label{sec:conclusion}
We have presented a new probabilistic AMP-based reconstruction algorithm for compressive 
imaging based on the cosparse analysis model with dual variables. This approach is similar to a finite temperature version of the GrAMPA algorithm, which makes it particularly adapted to 
natural images. Indeed, for this class of signals, the proposed algorithm $\scampi$ reaches state
-of-the-art perfomance, overcoming the previous best probabilistic methods. Furthermore, it is 
highly robust to noise channel uncertainty and does not require near-perfect knowledge to 
reach optimal performance. However, our experiments show that for cartoon images, 
consisting of only constant regions separated by instantaneous transitions, the zero 
temperature model does a better job of enforcing the piecewise-continuous prior on the image.

We have also shown through intensive numerical experiments that when we push the convex 
optimization TV-AL3 algorithm to its limits via oracle tuning parameter selection, it gives results very close to the proposed, easier to tune, approach. 
The fundamental reason for this empirical finding is yet to be understood.  
\section*{Acknowledgments}
This work has been supported in part by the ERC under the European Union's 7th Framework Programme Grant Agreement 307087-SPARCS, by the French Minist\`ere de la d\'efense/DGA and by the Swiss National Science Foundation grant number 200021-156672.
%
% \newpage
%
\section*{References}
\bibliographystyle{iopart-num.bst}
\bibliography{refs}
\end{document}